\documentclass[aps,prb,twocolumn,superscriptaddress,showpacs]{revtex4-1}
\usepackage{amsmath,amssymb,graphicx}

\usepackage[T1]{fontenc}
\usepackage{chancery}

\usepackage{xcolor}
\usepackage{physics}
\usepackage[bottom]{footmisc}

\IfFileExists{newtxtext.sty}   {\usepackage{newtxtext,newtxmath}}
{\IfFileExists{stix.sty}
	{\usepackage{stix}}
	{\IfFileExists{mathptmx.sty}
		{\usepackage{mathptmx}}{} } }

\usepackage{textcomp}

\usepackage{bm}

\IfFileExists{siunitx.sty}{\usepackage{booktabs,siunitx}}{}

\usepackage{color}
\definecolor{LinkColor}{rgb}{0.256,0.439,0.588}
\usepackage{hyperref}
\hypersetup{
	pdfauthor={Tsung-Cheng Lu},
	colorlinks=true,
	citecolor=LinkColor,
	linkcolor=LinkColor,
	urlcolor=LinkColor
}

\usepackage{pifont}
\usepackage{epstopdf}
\usepackage{caption}
\usepackage{subcaption}
\renewcommand{\raggedright}{\leftskip=0pt \rightskip=0pt plus 0cm}
\let\vec\mathbf
\newcommand{\beq}{\begin{eqnarray}}
\newcommand{\eeq}{\end{eqnarray}}

\begin{document}
	
    \title{Detecting Topological Order at Finite Temperature Using Entanglement Negativity}

	\author{Tsung-Cheng Lu} 
\affiliation{Department of Physics, University of California at San
	Diego, La Jolla, CA 92093, USA}
\author{Timothy H. Hsieh}
\affiliation{Perimeter Institute for Theoretical Physics, Waterloo, Ontario N2L 2Y5, Canada}
	\author{Tarun Grover}
\affiliation{Department of Physics, University of California at San
		Diego, La Jolla, CA 92093, USA}
	
	\begin{abstract}
We propose a diagnostic for finite temperature topological order using `topological entanglement negativity', the long-range component of a mixed-state entanglement measure. As a demonstration, we study the toric code model in $d$ spatial dimension for $d$=2,3,4, and find that when topological order survives thermal fluctuations, it possesses a non-zero topological entanglement negativity, whose value is equal to the topological entanglement entropy at zero temperature.  Furthermore, we show that the Gibbs state of 2D and 3D toric code at any non-zero temperature, and that of 4D toric code above a certain critical temperature, can be expressed as a convex combination of short-range entangled pure states, consistent with the absence of topological order.
	\end{abstract}
	
	
	\maketitle

A strongly interacting quantum many-body system at zero temperature can exhibit exotic order beyond Laudau-Ginzburg paradigm, dubbed topological order, whose defining property is that the ground state degeneracy depends on the topology of the space\cite{wen1989vacuum,wen1990ground,wen1990topological}. While the theory of topological order in ground states (i.e., zero temperature) is well developed, our understanding for topological order at finite temperature is less clear. In particular, the pursuit of a model supporting a stable topological order at finite temperature has been difficult since typically topological order is fragile against thermal fluctuations\cite{nussinov2008autocorrelations, nussinov2009sufficient,bravyi2009no_go,Poulin2013,brown2016review}. The most well-known model exhibiting finite-T topological order is the toric code model in four spatial dimensions\cite{dennis2002,alicki2010_4d}, and it remains unclear whether such a model exists below four dimensions. Apart from being a fundamental question in many-body physics, a stable finite-T topological order also has profound implications for quantum computing since it serves as a stable self-correcting quantum memory (encoded information is protected against thermal decoherence)\cite{dennis2002,yoshida2011}.

Even for models that support finite-T topological order, it is not obvious how to define an appropriate non-local order parameter. Hastings defined topological order at finite-T by the requirement that the corresponding thermal density matrix cannot be connected to a separable mixed state via a finite-depth quantum channel\cite{hastings2011}. While this provides a precise operational definition, it is still desirable to have a computable order parameter for finite-T topological order, analogous to the characterization of ground state topological order using topological entanglement entropy \cite{hamma2005bipartite, levin2006detecting, Kitaev06_1}.

Previous works have studied the subleading term $S_{\text{topo}}$ of the von Neumann entropy $S = - \tr \rho \log \rho$ at finite temperature, in models that support topological order at $T = 0$ \cite{castelnovo2007classical,castelnovo2007entanglement,castelnovo2008topological, Melko_Hubbard, hamma2012topological, swingle2016, li2019finite}. Nevertheless, $S_{\text{topo}}$ cannot distinguish quantum correlations from the classical ones: even a \textit{purely classical} $\mathbb{Z}_2$ gauge theory in three dimension has a non-zero $S_{\text{topo}}$ consistent with the fact that it exhibits  a self-correcting classical memory\cite{hasting_z2, yoshida2011}. 

\begin{figure}
	\centering
	\begin{subfigure}[b]{0.26\textwidth}
		\includegraphics[width=\textwidth]{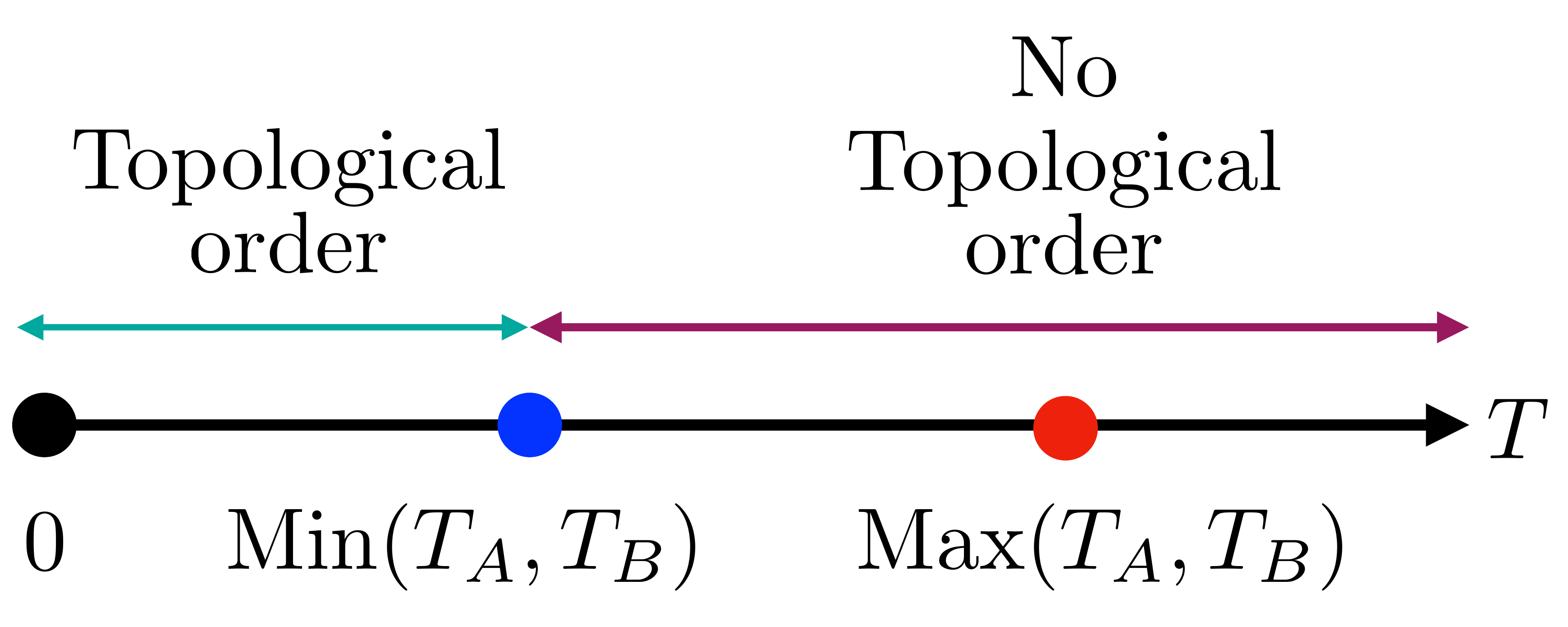}
		\label{fig:partition}
	\end{subfigure}
	\begin{subfigure}[b]{0.21\textwidth}
	\includegraphics[width=\textwidth]{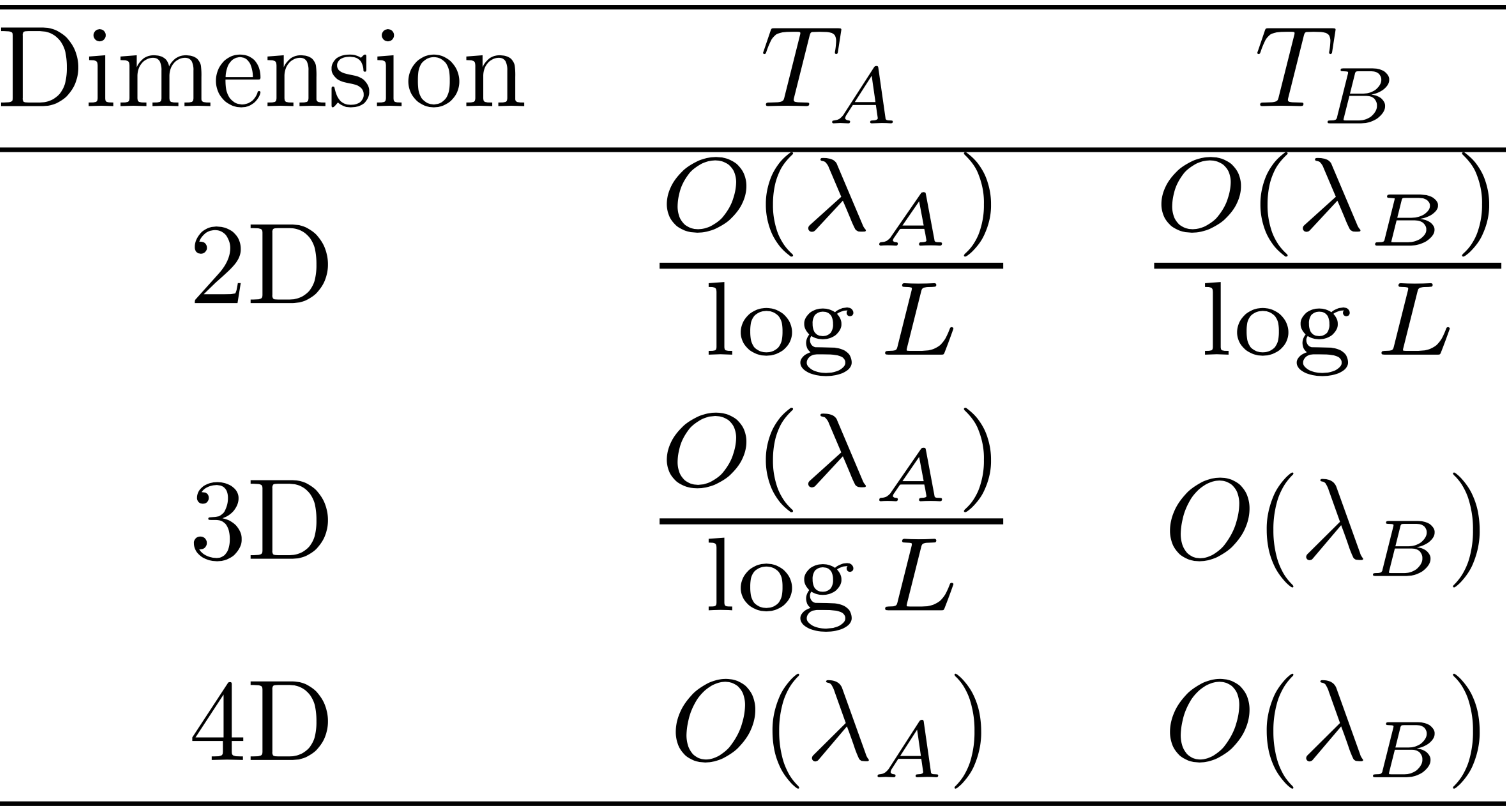}
	\label{fig:partition}
\end{subfigure}
	\begin{subfigure}[b]{0.45\textwidth}
		\includegraphics[width=\textwidth]{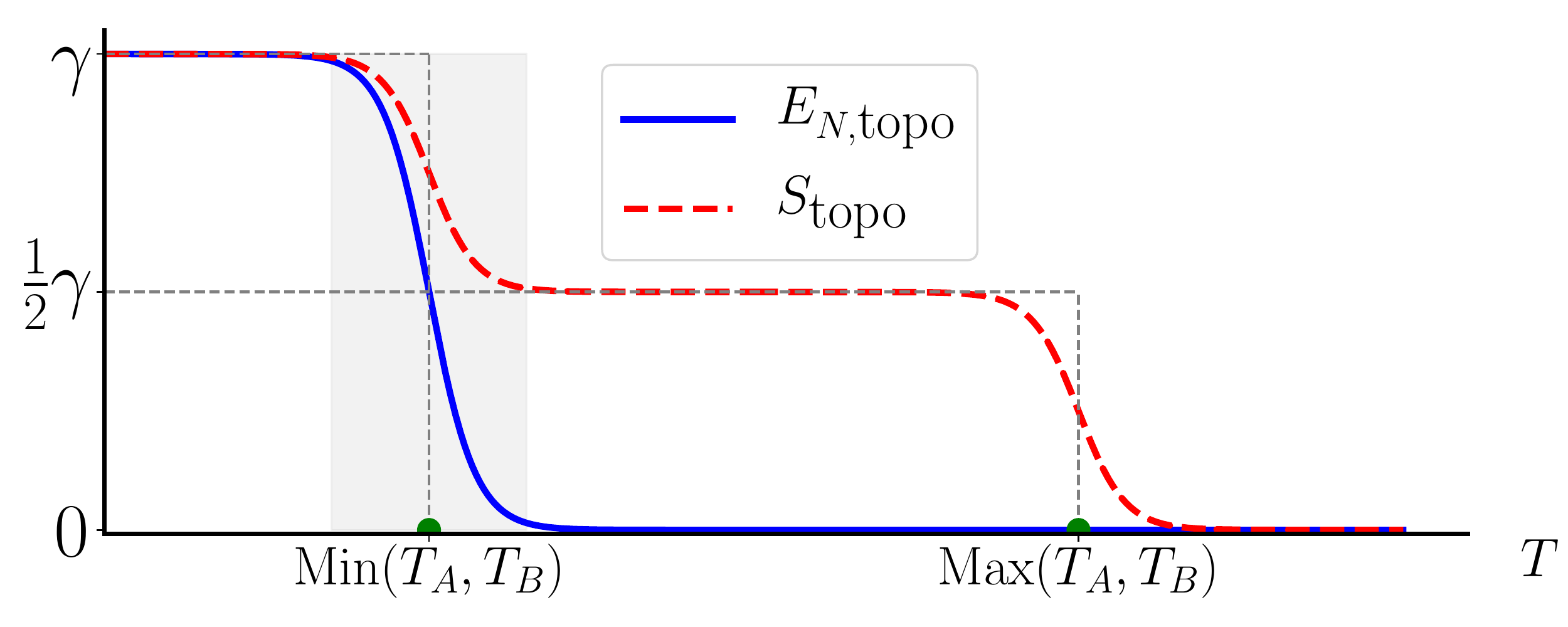}
	\end{subfigure}
	\caption{Upper panel: phase diagram of toric code models, where the critical temperatures $T_A$ and $T_B$ corresponding to the proliferation of two types of excitations depend on the spatial dimension. Lower panel: comparision between topological entanglement negativity $E_{N,\text{topo}}$ and topological von Neumann entropy $S_{\text{topo}}$ in toric code models of size $L$. As $L\to \infty$, $E_{N,\text{topo}}=0$ for $T>\text{Min}(T_A,T_B)$, consistent with the absence of topological order while $S_{\text{topo}}$ remains non-zero in the regime $\text{Min}(T_A,T_B)< T< \text{Max}(T_A,T_B)$. When $\text{Min}(T_A,T_B) \neq 0$ as $L \to \infty$, the behavior of $E_{N,\text{topo}}$ shown close to the critical point (= the shaded region) is just a schematic and we do not probe that region.}  \label{fig:alld}
\end{figure}

In this paper, we propose an entanglement-based diagnostic for finite-T topological order that is sensitive only to quantum correlations. Specifically, we employ entanglement negativity $E_N$, a mixed-state entanglement measure, to quantify non-local quantum correlations resulting from finite-T topological order. The intuition behind our approach is that if a mixed state possesses long-range entanglement, then it is non-separable over a length scale proportional to the system size, and therefore, such entanglement cannot be undone via any finite-depth quantum channel. 

Given a Gibbs state corresponding to a local model, for a smooth entangling boundary, one can express $E_N$ as a sum of local and non-local terms\cite{lu2019structure}, analogous to the case of entanglement entropy for gapped ground states\cite{grover2011entanglement}:  $E_N=E_{N,\text{local}}+E_{N,\text{topo}}$, where $E_{N,\text{local}}=\alpha_{d-1}L_A^{d-1} +  \alpha_{d-3}L_A^{d-3} +\cdots$ characterizes the short-range entanglement, while $E_{N,\text{topo}}$ denotes the non-local entanglement, which is not expressible as a functional of local curvature along the entangling boundary. We will denote  the non-local term as `topological entanglement negativity' and use it as a diagnostic for finite-T topological order.

 We will primarily focus on toric code models at finite-T in $d$ spatial dimensions for $d=2,3,4$. A d-dimensional toric code Hamiltonian can be written as $H= -\lambda_A\sum_s A_s   - \lambda_B \sum_p B_p$ in which $A_s$ are products of Pauli-X operators, and $B_p$ are products of Pauli-Z operators (their precise forms depend on the dimensionality). This model can be thought of as a sum of two classical gauge theories in two different bases, which has an interesting consequence: the partition function factorizes, $Z = Z_A Z_B/2^N$ where $Z_A =  \tr\left( e^{ \beta \lambda_A \sum_s A_s}\right)$, $Z_B =  \tr\left( e^{ \beta \lambda_B \sum_p B_p}\right)$, and $N$ is the number of spins.  Due to this structure, the toric code model has two critical temperatures $T_A$ and $T_B$ above which the excitations corresponding to  $A_s$ and $B_p$ operators proliferate. In the language of the gauge theory, these temperatures correspond to confinement-deconfinement transition for Wilson operators $W_x$ and $W_z$ respectively, where $W_x/W_z$ is a product of connected $A_s/B_p$ operators and is a gauge invariant under the local gauge transformation generated by $B_p/A_s$. Intuitively, at a finite temperature, a stable topological order can protect the encoded qubits against the thermal decoherence without the need of any active error correction, only when both types of excitations are suppressed, that is, below Min($T_A,T_B$). On the other hand, if only one type of excitations is suppressed, i.e. in the temperature regime Min($T_A,T_B$) and Max($T_A,T_B$), the other type of excitation destroys the topological order, and the model can only realize a self-correcting \textit{classical} memory\cite{yoshida2011,castelnovo2007entanglement,castelnovo2008topological,hasting_z2}.

Our main result is summarized in Fig.\ref{fig:alld}. Through an explicit calculation, we find that topological entanglement negativity is nonzero only when the temperature is simultaneously below both critical temperatures associated with the proliferation of two types of excitations, in line with the aforementioned heuristics. In strong contrast, $S_{\text{topo}}$ remains nonzero (drops to half of its ground state value) when temperature is between the lower and upper critical temperatures \cite{castelnovo2007entanglement,castelnovo2008topological}. 

\textbf{\textit{Disentangling toric codes at finite-T}}--- Before discussing topological entanglement negativity for toric code models in detail, here we provide intuition for finite-T topological order by decomposing a mixed state of interest into a convex sum of pure states: $\rho = \sum_i p_i |\psi_i\rangle \langle \psi_i|$. If each $|\psi_i\rangle $ is short-range entangled, then preparing $\rho$  requires only the ability to generate the probability distribution $\{p_i\}$, and constructing short-ranged entangled states $|\psi_i\rangle$, tasks which can be done with resources that do not scale with the system size. Alternatively, one can purify $\rho$ to obtain a state that can be constructed with a finite depth unitary (\cite{hastings2011}, see also \cite{supplement} for an explicit construction for toric code). One hint for such a decomposition comes from `minimally entangled typical thermal state' (METTS) ansatz\cite{white_metts}: $\rho= e^{-\beta H}/Z= \sum_m p_m |\phi_m\rangle \langle \phi_m|$ where each $\ket{\phi_m}$ is a METTS obtained from imaginary time evolution of a product state $\ket{m}$: $|\phi_m\rangle \sim e^{-\beta H/2} |m\rangle$. $p_m=\bra{m}e^{-\beta H} \ket{m} /Z$ is the probability corresponding to $\ket{\phi_m}$. Using such decomposition, we now show that the Gibbs state of the toric code model in arbitrary spatial dimension is not topologically ordered above Min($T_A,T_B$). 

First consider METTS obtained from a product state $\ket{m}$ in the Z basis: $\ket{\phi_m(T)}\sim e^{\beta/2\sum_{s}A_s} e^{\beta/2\sum_{p}B_p}  \ket{m}  \sim  e^{\beta/2\sum_{s}A_s}\ket{m}$. One finds all such METTS $\ket{  \phi_m(T)}$ at temperature $T>T_A$ are short-range entangled since they can be adiabatically connected to the infinite temperature METTS $\ket{\phi_m(T\to \infty)  }$, i.e. a product state, without encountering a phase transition/critical point. Therefore $\rho$ is not topologically ordered for $T>T_A$. Similarly, one can also decompose $\rho$ using METTS obtained by imaginary time evolving the product state in X basis to deduce that $\rho$ is not topologically ordered for $T>T_B$. Combining these two observations proves the absence of topological order in toric code for temperature $T> \min (T_A,T_B)$. Note that this result applies to all CSS code Hamiltonians $H=-\lambda_A\sum_{i}S_i^{(X)}-\lambda_B\sum_{i}S_i^{(Z)}$\cite{css_Steane,css_Shor}, where each local commuting term $S_i^{(X/Z)}$ is a product of Pauli-X/Z operators. Using this result and the observation in Ref.\cite{haah2013lattice, bravyi2013, finite_T_xcube,zohar}, one immediately proves the absence of finite-T topological order in the more exotic topological models such as X-cube model\cite{xcube}, a type-I fracton model, and Haah's code\cite{Haah}, a type-II fracton model. As an aside, each METTS $\ket{\phi_m(T)}$ is the ground state of a local parent Hamiltonian\cite{supplement}, which can be explicitly constructed using an approach analogous to Ref.\cite{PhysRevB.93.205159, topo_scar}.

\begin{figure}
	\centering
	\includegraphics[width=0.5\textwidth]{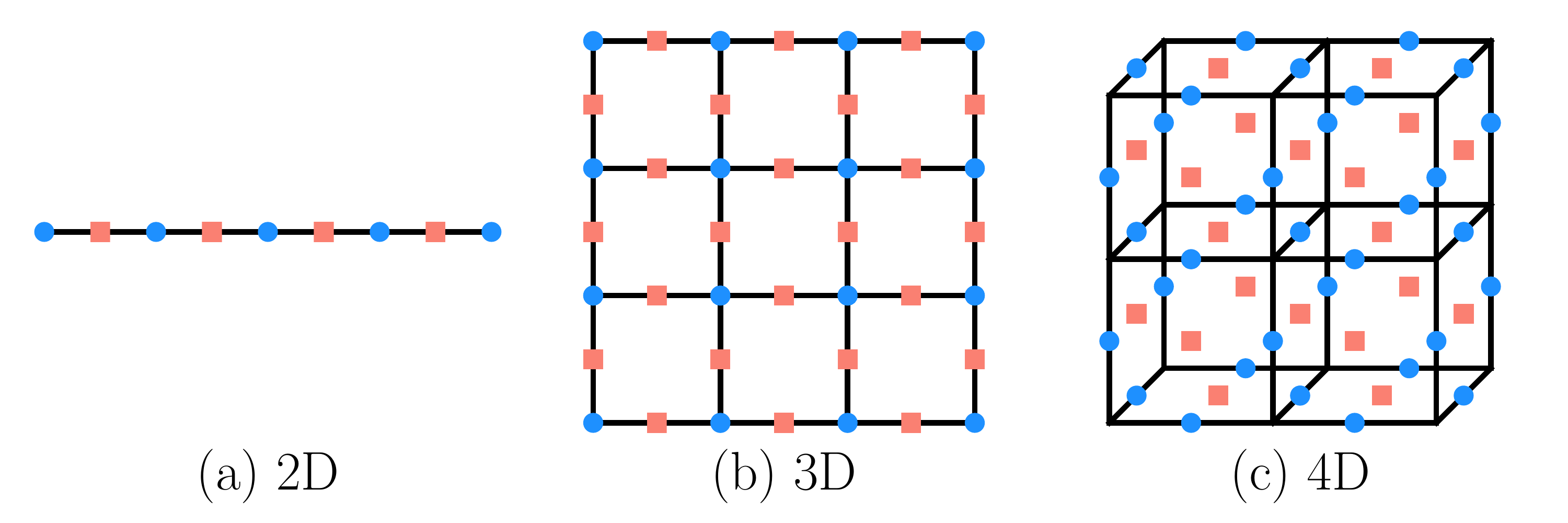}
	\caption{The boundary operators in toric code for various spatial dimensions, where blue circles and red squares label $A_i$ and $B_j$ operators respectively. (a) 1D bipartition boundary in 2D toric code, where $A_i$ live on sites, and $B_j$ live on links. (b) 2D bipartition boundary in 3D toric code, where $A_i$ live on sites, and $B_j$ live on links. (c) 3D bipartition boundary in 4D toric code, where $A_i$ live on links, and $B_j$ live on faces.}
	\label{fig:boundary}
\end{figure}

\begin{figure}
	\centering
	\includegraphics[width=0.5\textwidth]{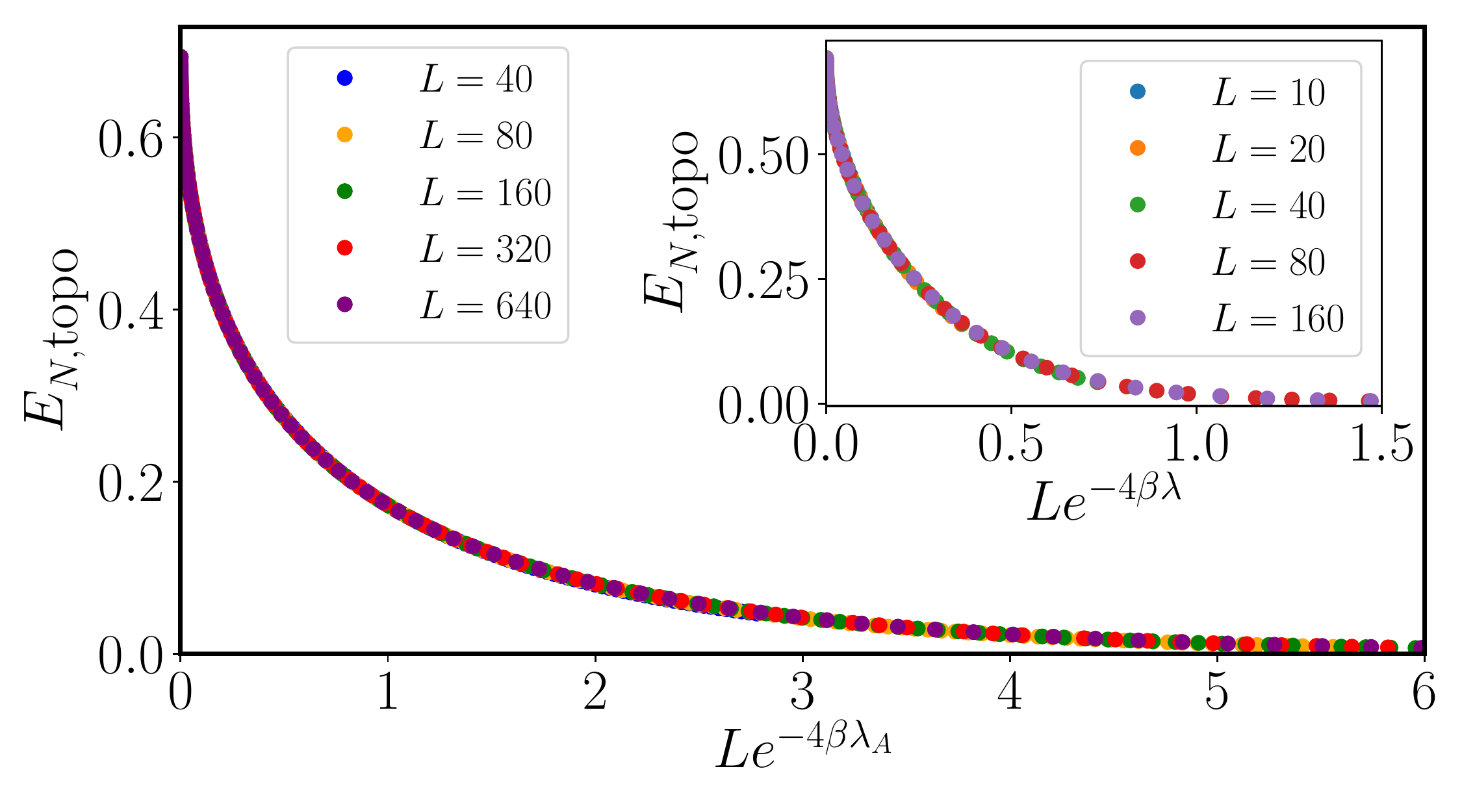}
	\caption{Scaling collapse of topological negativity in 2D toric code as $\lambda_B\to\infty$ (Eq.\ref{2D:topo_negativity}). $L$ is the size of the bipartition boundary, $\beta$ is the inverse temperature, and $\lambda_A$ is the coefficient of the star operators $A_s$. Inset: Scaling collapse of topological negativity in 2D toric code at $\lambda=\lambda_A=\lambda_B$ using classical Monte Carlo combined with transfer matrix method.}
	\label{fig:2d_nega}
\end{figure}
\textbf{\textit{General scheme for calculating negativity}}--- The above calculation using METTS ansatz shows when a state is \textit{not} topologically ordered. To understand the fate of topological order for $T <$ Min($T_A,T_B$), we now turn to characterizing the mixed state entanglement of the Gibbs state using entanglement negativity \cite{eisert99, vidal2002, plenio2005logarithmic}. Given a density matrix $\rho$ acting on the Hilbert space $\mathcal{H}_A\otimes   \mathcal{H}_B$: $\rho=\sum_{A,B,A',B'}  \rho_{AB,A'B'}\ket{A}\ket{B}\bra{A'}\bra{B'}$, one defines a partially transposed matrix $\rho^{T_B}$ as: $\rho^{T_B}=\sum_{A,B,A',B'}  \rho_{AB,A'B'}\ket{A}\ket{B'}\bra{A'}\bra{B}$. The negativity is defined as $E_N=\log \norm{\rho^{T_B}}_1= \log \left(  \sum_i \abs{\lambda_i}  \right)$
where $\lambda_i$ are the eigenvalues of $\rho^{T_B}$. Relatedly, we define $n$-th Renyi negativity $R_n$: $R_n= b_n  \log \left(   \frac{  \tr{  \left(\rho^{T_B} \right)^n   }  }{ \tr\rho^n  }      \right)$ where $b_n=\frac{1}{1-n}$ for odd $n$ and  $\frac{1}{2-n}$ for even $n$. The prefactor is chosen so that when $\rho$ is a pure state, $R_n$ reduces to Renyi entanglement entropy: $R_n=S_n,~S_{n/2}$ for odd $n$ and even $n$ respectively. Further, negativity  $E_N=\lim_{n\to 1}R_n$, assuming $n$ is even. We now turn to study the negativity of toric code in $d=2,3,4$ dimension.

The negativity in 2D toric code at finite temperature was discussed in Ref.\cite{castelnovo2018}, which focuses on how finite temperature excitations decrease quantum correlations, eventually leading to vanishing of negativity above a `sudden death temperature' $T_d > 0$. Here, we instead focus on the topological part $E_{N,\text{topo}}$, which captures the topological order.

We now present an approach motivated by Ref.\cite{lu2018singularity} to taking the partial transpose of Gibbs states for stabilizer code Hamiltonians $H=-\sum_{m}S_m$, where $S_m$ is a product of Pauli matrices over sites with $S_m^2=1$.  Using $e^{\beta S_m}=\cosh \beta + S_m \sinh\beta$, one can expand the Gibbs state as $e^{-\beta H}\sim \sum_{\{ x_m\}} \prod_m ( S_m \tanh\beta   )^{x_m}$, where $x_m=0/1$ indicates the absence/presence of $S_m$. Consider a subregion $\mathcal{R}$ and its complement $\overline{\mathcal{R}}$, taking partial transpose over $\overline{\mathcal{R}}$ in computational bases gives $ \left(\prod_{m}S_m^{x_m}  \right)^{T_{\overline{\mathcal{R}}}}   = \psi(\{x_m\}) \prod_{m}S^{x_m}_m$, where the sign $\psi(\{x_m \})=1/-1$ corresponds to even/odd number of Pauli$-Y$s in $\prod_m S_m^{x_m}$ on region $\overline{\mathcal{R}}$. Since stabilizers $S_m$ supported only in $\mathcal{R}$ or $\overline{\mathcal{R}}$ always give even number of Pauli$-Y$s in $\overline{\mathcal{R}}$ for toric codes, the sign $\psi$ is solely determined by the appearance of the stabilizers across the bipartition boundary. This implies the partial transpose only acts on the bipartition boundary part of the Gibbs state: $\rho^{T_{\overline{\mathcal{R}}}} =\frac{1}{Z} \left( e^{-\beta H_{\mathcal{R}\overline{\mathcal{R}}}}  \right)^{T_{ \overline{\mathcal{R}}   }}  e^{-\beta  (H_\mathcal{R}+H_{\overline{\mathcal{R}}} )}$, where $H_\mathcal{R}/H_{\overline{\mathcal{R}}}$ denotes the part of $H$ supported on $\mathcal{R}/\overline{\mathcal{R}}$, and $H_{\mathcal{R}\overline{\mathcal{R}}}$ denotes the interaction between $\mathcal{R}$ and $\overline{\mathcal{R}}$. Define $\{A_i\}$, $\{B_j\}$ as the star and plaquette operators across the bipartition boundary, one finds

\begin{equation}\label{eq:1}
e^{-\beta H_{\mathcal{R} \overline{\mathcal{R}}}}  \propto \sum_{  \substack{ \{n_i=0,1\}   \\ \{\sigma_j=0,1\} }  }  \prod_{i=1}^{N_s^{\partial}} (A_i\tanh(\beta \lambda_A))^{n_i} \prod_{j=1}^{N_p^{\partial}}  (B_j \tanh(\beta \lambda_B  )  )^{\sigma_j}.
\end{equation}
As mentioned above, taking partial transpose on a Pauli string introduces a sign determined by the number parity of Pauli$-Y$s in region $\overline{\mathcal{R}}$: $\left\{  \left[  \prod_{i=1}^{N_{s}^{\partial}   }    A_{i}^{n_i}   \right]    \left[  \prod_{ j=1 }^{N_p^{\partial}} B_{j}^{\sigma_{j}}   \right] \right\}^{T_{ \overline{\mathcal{R}}}}   =  \left[  \prod_{i=1}^{N_{s}^{\partial}   }      A_{i}^{n_i}   \right]    \left[  \prod_{ j =1}^{N_p^{\partial}}   B_{j}^{\sigma_{j}}   \right]  \psi(   \{ n_i \}  ,  \{ \sigma_j \})$. Since Pauli$-Y$s only occur from products of Pauli$-X$s and Pauli$-Z$s from neighboring star and plaquette operators, we find $\psi(   \{ n_i \}  ,  \{ \sigma_j \}) =\prod_{j=1}^{N_{p}^{\partial}}  \left(   \prod_{  i \in \partial j  }  \tau_i \right)^{ \sigma_{j}   }$ where we have introduced the Ising variables $\tau_i=1-2n_i \in  \{\pm 1 \}$. One can now sum over the $\sigma_j$ variables and express $\left(  e^{-\beta H_{\mathcal{R} \overline{\mathcal{R}}}}   \right)^{T_{ \overline{\mathcal{R}}}}$ as a partition function over $\tau_i$: $\left(  e^{-\beta H_{\mathcal{R} \overline{\mathcal{R}}}}   \right)^{T_{ \overline{\mathcal{R}}}} 
=\cosh[N_s^{\partial}](\beta\lambda_A)  \sum_{   \{ \tau_i  \}   }  e^{ -H'  (   \{ \tau_i  \}   , \{A_i\} ,\{B_j\}   )}$, where 
\begin{equation} \label{Heff}
-H' =  \sum_{i=1}^{N_s^{\partial}}     \frac{1-\tau_i}{2} \log \left(  A_{i} \tanh(\beta \lambda_A)\right)   +  \beta \lambda_B  \sum^{N_p^{\partial}}_{j=1}  B_j   \prod_{i\in \partial j    }  \tau_i
\end{equation}

Replacing the commuting operators $A_s,B_p$ with $\pm 1$ gives the eigenspectrum of $\rho^{T_{\overline{\mathcal{R}}}}$. Interestingly, we find that positive/negative signs of eigenvalues reflect the parity of braids between star and plaquette operators on the bipartition boundary at zero temperature\cite{supplement}. This formalism can be used to derive the partial transpose of a Gibbs state in other stabilizer models, such as the two dimensional Wen plaquette model\cite{wen2003quantum} or the more exotic fracton models\cite{xcube,Haah}. We provide an alternative derivation of the above result using the matrix product state representation\cite{supplement}.

\textbf{\textit{2D toric code}}---  Recall that the star term $A_s(=\prod_{i\in s} X_i)$ is the tensor product of four-Pauli $X$ operators on the star labeled by $s$, and $B_p(=\prod_{i\in p } Z_i)$ is the tensor product of four-Pauli $Z$ operators on the plaquette labeled by $p$. Defining the model on a two-torus gives four-fold degenerate ground states, where two qubits can be encoded, and are immune from local perturbation. However, both types of excitations (by flipping the sign of $A_s$ and $B_p$) are point-like charges which proliferate at any non-zero temperature to destroy the topological order and the encoded quantum information in the ground subspace. The topological entanglement negativity is $\log(2)$ at $T = 0$ \cite{castelnovo2013, vidal2013},  and here we show that the absence of the topological order at finite temperature can be captured by the absence of the topological (Renyi) negativity.

In this case, the bipartition boundary is a 1D system of length $L$ consisting of $L$ star and $L$ plaquette operators (see Fig.\ref{fig:boundary}a). Eq.\ref{Heff} implies that $H'$ corresponds to a 1D Ising model in a magnetic field, which yields negativity $E_N = \log \expval{ \abs{Z\left( \{A_i\} ,\{B_j\} \right)}  }$. $Z(\{A_i  \},   \{B_j\})=\frac{1}{ \left[ \cosh\beta \lambda_B\right]^L} \sum_{ \{ \tau_i=\pm 1\}} e^{-H' (\{\tau_i  \}  , \{  A_i \} , \{B_j  \}   ) }$ with the angled brackets denoting the `disorder average' over the variables $\{A_i=\pm1 \}$ and $\{B_j\pm1 \}$. This expression was first obtained in Ref.\cite{castelnovo2018} using a replica trick, and we have provided an alternative derivation.

We first consider the limit $\lambda_B\to \infty$, forbidding magnetic charges in the $\mathbb{Z}_2$ gauge theory. Since in this limit the system realizes only a self-correcting \textit{classical} memory instead of a quantum memory, it is a good starting point to see if the topological entanglement negativity is insensitive to long-distance classical correlations. Considering two connected regions separated by a closed boundary of size $L$, and defining $x=\tanh(\beta \lambda_A)$, one finds a compact expression for the negativity $E_N=\alpha L -  E_{N,\text{topo}}$. The area-law coefficient $\alpha=\log \left( 1+x \right)$ was first derived in Ref.\cite{castelnovo2018}, and here we instead focus on the topological entanglement negativity\cite{supplement}:
\begin{widetext}
\begin{equation}\label{2D:topo_negativity}
  E_{N,\text{topo}}= - \log   \left\{     \frac{1}{2}    +   \frac{1}{2} \left(  x^{1/2}+x^{-1/2} \right)^{-L}    \binom{L}{\frac{L}{2}+1}    \left[ \frac{1}{x} ~  _2F_1( 1,-\frac{L}{2}+1 ;  \frac{L}{2}+2 ;-\frac{1}{x}  )   -  x  ~ _2F_1( 1,-\frac{L}{2}+1 ;  \frac{L}{2}+2 ;-x )       \right]     \right\}
\end{equation}
\end{widetext}
where $_2F_1$ is the hypergeometric function. While $E_{N,\text{topo}}$ is $\log 2$ at zero temperature \cite{vidal2013, castelnovo2013, wen2016edge, wen2016topological},  it is exactly zero for any finite temperature as $L\to \infty$. Interestingly, for a finite $L$ at low temperature, one finds $E_{N,\text{topo}}$ only depends on the scaling variable $Le^{-4\beta\lambda_A}$, as one may also verify by an asymptotic expansion of the hypergeometric function (see Fig.\ref{fig:2d_nega}). This scaling variable represents the number of pairs of anyons thermally excited on the boundary.  We also obtain analytical expressions for all even and odd Renyi negativities and find they depend respectively on $Le^{-4 \beta \lambda_A}$ and $Le^{-2 \beta \lambda_A}$\cite{supplement}. Next, for general $\lambda_A$ and $\lambda_B$, we combine a classical Monte Carlo method and a transfer matrix method to calculate negativity, and find qualitatively same behavior as in the limit $\lambda_B \to \infty$ (see Fig.\ref{fig:2d_nega} inset).   We also use a generalized transfer matrix method to analytically show that the topological (Renyi) negativity is $\log2$ at zero temperature and vanishes for any finite temperature as $L\to \infty$\cite{supplement}.

\textbf{\textit{3D toric code}}--- Here the star operator $A_s(=\prod_{i\in s} X_i)$ is the product of six Pauli-X operators on the links emanating from a vertex of the cubic lattice, and the plaquette operator $B_p(=\prod_{i\in p} Z_p)$ is the product of four Pauli-Z operators on the links of a plaquette. Choosing $\lambda_A, \lambda_B>0$, the ground subspace is specified by $A_s=B_p=1$, and imposing periodic boundary conditions results in 8 orthogonal ground states, where one can encode 3 qubits. While flipping the sign of $B_p$ gives loop-like excitations, which is suppressed below the critical temperature corresponding to the 3D $Z_2$ gauge theory confinement transition, flipping $A_s$ gives point-like excitations, which proliferate at any nonzero temperature to destroy the topological order. Now we will show that topological negativity can again diagnose finite-T topological order.

Given a bipartition boundary of linear size $L$, there are  $L^2$ boundary star operators $A_i$ living on the lattice sites of the two dimensional boundary and $2L^2$ plaquette operators $B_{ij}$ living on the links $\expval{ij}$ (Fig.\ref{fig:boundary}b). We again utilize the general formalism (Eq.\ref{Heff}) specialized in this geometry to calculate negativity.

To separately see the effects of point like versus loop like excitations, we first consider $\lambda_B\to \infty $ to prohibit loop-like excitations. In this limit, negativity is exactly the same as the one in 2D by taking $L\to L^2$\cite{supplement}, indicating the presence of only point-like excitations. The topological negativity is exactly given by Eq.\ref{2D:topo_negativity} by taking $L\to L^2$, and hence it vanishes in the thermodynamic limit at any non-zero temperature.

In contrast, taking $\lambda_A\to \infty$ prohibits point-like excitations, and thus the finite-T topological order exists up to the critical temperature of 3+1 D $Z_2$ gauge theory, above which the loop-like excitations proliferate. Despite Eq.\ref{Heff} giving the analytical form of $\rho^{T_{\overline{\mathcal{R}}}}$, the calculation of negativity is challenging because (1) each eigenvalue of $\rho^{T_{\overline{\mathcal{R}}}}$ requires calculating the partition function of 2D Ising model of $\tau_i$ spins subject to arbitrary given $B_{ij}$ and $A_i$ (2) plaquette operators $B_p$ cannot be freely chosen for the eigenspectrum due to the local constraint $\prod_{p\in \partial \text{cubic}}B_p=1$. Therefore, we turn to Renyi negativity $R_n$, which can be shown as the free energy difference of two statistical mechanics models: $R_n \sim   \log \tilde{Z} -\log Z $, where $Z$ is a partition function of 3D $Z_2$ gauge theory, $\tilde{Z}$ is a partition function of 3D $Z_2$ gauge theory coupled to $n$ replicas of 2D Ising models. The absence of zero temperature critical point (due to $\lambda_A \rightarrow \infty$) allows one to perform a low temperature perturbative calculation for Renyi negativity $R_n$ and find\cite{supplement}

\begin{equation}
R_n=\alpha L^2- R_{n,\text{topo}}
\end{equation}
where 
\begin{equation}
\alpha =\begin{cases}
\log 2-\frac{1}{n-2} \left[ \binom{n}{2} e^{-16\beta\lambda_B}  + 2  \binom{n}{2} e^{-24\beta\lambda_B}  +.. \right]~\text{for even } n\\
\log 2 -\frac{n}{n-1}   \left[ e^{- 8\beta \lambda_B}+2e^{-12\beta \lambda_B} + \frac{9}{2} e^{-16\beta \lambda_B} +.. \right] ~\text{for odd } n
\end{cases}
\end{equation}
and $R_{n,\text{topo}}= \log 2$. In fact, using the linked cluster theorem, which demands that only the excitations given by connected spin flips contribute to the logarithm of partition functions, we find those connected `diagrams' only contribute to the area law component of $R_n$ without changing  $R_{n,\text{topo}}$.  Hence, we expect $R_{n,\text{topo}}$ remains $\log 2$ until the breakdown of the perturbative series, which occurs at the critical point of the 3D $Z_2$ gauge theory. Since   $R_{n,\text{topo}}$ is independent of $n$, we conclude that $E_{N,\text{topo}} = \log(2)$ as well.

\textbf{\textit{4D toric code}}--- Finally we discuss the toric code in four spatial dimension, which realizes  finite-T topological order  \cite{dennis2002}. Spins reside on each face of the 4D hypercube, and the Hamiltonian reads
$H=-\lambda_A \sum_l A_l -\lambda_B \sum_{c}B_c$,
where $l$, $c$ label links and cubes. $A_l$ is the product of 6 Pauli-X operators on the faces adjacent to the link $l$, and $B_c$ is the product of 6 Pauli-Z operators on the faces around the boundary of the cube $c$. Flipping the sign of $A_l$ or $B_c$ gives loop-like excitations living on the boundary of two dimensional open membranes, whose energy scales with the loop size. Therefore this model has two finite temperature critical points, corresponding to the proliferation of two loop-like excitations, and it supports finite-T topological order up to temperature $ T_c \propto \text{Min}(\lambda_A, \lambda_B$). 

The boundary of the 4D hypercube is a 3D cubic lattice, where the boundary operators are $A_l$ living on every link and $B_f$ living on every face (Fig.\ref{fig:boundary}c). Using Eq.\ref{Heff}, we find that at zero temperature topological negativity is $2\log 2$, consistent with the topological entanglement entropy\cite{grover2011entanglement}. We perform a low temperature perturbative calculation for Renyi negativity of even $n$, and find $R_n= \alpha L^3 -R_{n,\text{topo}}$ where $\alpha=   2\log 2 -\frac{3n(n-1)}{2(n-2)}\left(  e^{-16\beta \lambda_A} +e^{-16\beta \lambda_B}  \right)    + \cdots $, 
and the topological part $R_{n,\text{topo}}$ remains the ground state value $2\log 2$ \cite{grover2011entanglement}. Similar to the 3D toric code in $\lambda_A\to \infty $ limit, we expect $R_{n,\text{topo}}$ remains unchanged up to $T_c$.

\textbf{\textit{Summary}}---In this work we propose the topological entanglement negativity as a diagnosis for finite-temperature topological order and correspondingly, a self-correcting quantum memory. We find it successfully detects the absence of finite-T topological order in 2D toric code.  We demonstrated the robustness of topological entanglement in 3D toric code when the point-like excitations are suppressed, and in the 4D toric code, consistent with finite-T topological order.  Using METTS ansatz, we also provided an explicit decomposition of the Gibbs state in terms of short-range entangled pure states above Min($T_A,T_B$) where $T_A, T_B$ are defined in Fig.\ref{fig:alld}.

One application of our proposal is to disentangle quantum correlations from classical ones in realistic models relevant to frustrated magnets. For example, spin-ice systems exhibit emergent photons and monopoles below the degeneracy temperature of classical configurations \cite{castelnovo2008magnetic}, irrespective of whether the ground state is topological ordered or not. Negativity provides a precise diagnostic that distinguishes systems where the degenerate states coherently superpose to yield a topologically ordered state \cite{hermele2004}  (`quantum spin ice'), from systems that exhibit only classical emergent electromagnetism.


An important question remains: what is the critical behavior of topological entanglement negativity across a finite critical point, above which quantum memory is lost? It has been shown that the thermodynamic criticality of the 4D toric code follows the 4D Ising universality\cite{zohar}. However, the transition is intrinsically `quantum mechanical' since it is associated with the loss of universal, long-distance quantum correlations as well, and studying the critical behavior of topological entanglement negativity may provide new insights for such a finite temperature `quantum phase transition'.

 \emph{Acknowledgments--}
The authors thank John McGreevy for helpful discussions, and Matt Hastings for useful comments on the draft.
This research was supported by the Natural Sciences and Engineering Research Council of Canada (NSERC), the Canada Research Chair program, and the Perimeter Institute for Theoretical Physics. Research at Perimeter Institute is supported by the Government of Canada through Industry Canada and by the Province of Ontario through the Ministry of Research \& Innovation. TG is supported as an Alfred P. Sloan Research Fellow.

\bibliography{v1bib}
\renewcommand\refname{Reference}

\newpage 
\appendix

\onecolumngrid


\section{Decomposition of a thermal state using minimally entangled typical thermal states (METTS)}
One route to build intuition for topological order in a thermal state $\rho$ is to decompose $\rho$ using METTS. The central idea is to introduce a complete product state basis $1=\sum_m\ket{m}\bra{m}$, and break a thermal state with inverse temperature $\beta$ into two copies of thermal state with inverse temperature $\beta/2$:
\begin{equation}
	\rho= \frac{1}{Z} e^{-\beta H}  = \frac{1}{Z}e^{-\beta H/2} \sum_m \ket{m} \bra{m} e^{-\beta H/2} = \sum_m p_m\ket{\phi_m} \bra{\phi_m}.
\end{equation}
$\ket{\phi_m} \sim e^{-\beta H/2 } \ket{m}$ is the METTS, and the symbol $\sim$ refers to the equal sign up to a normalization constant. $p_m=\frac{\bra{m}e^{-\beta H} \ket{m}    }{Z}$ is the probability corresponding to $\ket{\phi_m}$, and $\sum_m p_m=1$. Choosing $H$ as the toric code Hamiltonian, one finds 
\begin{equation}
	\ket{\phi_m} \sim \prod_p \left(    1+B_P\tanh(\beta\lambda_B/2)  \right)    \prod_s \left(    1+A_s\tanh(\beta\lambda_A/2)       \right)    \ket{m}
\end{equation}

\subsection{2D toric code}
Choosing $\ket{m}$ as a product state in $X$ basis, the METTS becomes $
\ket{\phi_m} \sim \prod_p \left(    1+B_P\tanh(\beta\lambda_B/2)  \right)    \ket{m}$.
By expanding the product over plaquette $p$, whenever $B_p$ is chosen to create a closed loop on the boundary of $p$, the factor $ \tanh(\beta\lambda_B/2)$ follows. However, notice that a loop created by $B_p$ can also be obtained by $\prod_{p'\neq p}B_{p'}$ due to $\prod_p B_p=1$ on a two-torus. Thus

\begin{equation}
	\begin{split}
		\ket{\phi_m}  &\sim \sum_{C}  \left\{  \left[     \tanh(\beta\lambda_B/2) \right]^{ N_{\text{area}}(C)  } + \left[     \tanh(\beta\lambda_B/2) \right]^{N- N_{\text{area}}(C)  }              \right\}  \ket{C_m}  \\
		&=  \sum_{C}  \left[    e^{ -N_{\text{area}}(C)  \abs{  \log\left( \tanh(\beta \lambda_B/2)  \right) }   }   +  e^{ -  ( N- N_{\text{area}}(C) ) \abs{  \log\left( \tanh(\beta \lambda_B/2)  \right) }   }  \right] \ket{C_m},
	\end{split}
\end{equation}
where $\ket{C_m}$ is a classical loop state created from a reference state $\ket{m}$, $N$ is the total number of plaquettes, $N_{\text{area}} (C)$ is the area enclosed by the loop $C$. This expression implies the METTS cannot support arbitrary large closed loop configurations at any finite temperature, implying the absence of topological order. Since all $\ket{\phi_m}$ are short-range entangled, $\rho$ is not topological ordered at any fintie T.

Here we construct the exact parent Hamiltonian, for which $\ket{\phi_m}$ is the ground state. For simplicity, let's first consider $\ket{\phi_m}$ by choosing $\ket{m}$ as the product state with $+1$ eigenvalue of $X_l$ for all links $l$ . Define the operator
\begin{equation}
	Q_l =e^{-\beta \lambda_B\sum_p^{l \in \partial p}  B_p   }  -X_l,
\end{equation}
where $\sum_p^{l\in\partial p  }$ denotes the summation over $B_p$ containing the link $l$, we find 
\begin{equation}
	Q_l \ket{\phi_m}  \sim  \left[ e^{-\beta\lambda_B \sum_p^{l \in \partial p}  B_p   }  -X_l \right]  e^{\frac{1}{2}  \beta \lambda_B  \sum_p B_p } \ket{m}=0
\end{equation} 
for all $l$. A quick way to see the above equality is to note that the two terms in $Q_l$ play exactly the same role: providing a minus sign for the exponent of $e^{\frac{1}{2} \beta \lambda_B B_p}$ when $B_p$ contains the link $l$. Also, $Q_l$ is a positive semidefinite operator by observing $Q_l^2= 2Q_l  \cosh( \beta \lambda_B \sum_p^{l\in\partial p}  B_p   )  $. This implies $\ket{\phi_m}$ is the ground state with zero eigenenergy of the local Hamiltonian 
\begin{equation}
	\widetilde{H}=\sum_l  Q_l.
\end{equation}
Suppose that we start with a different $\ket{m}$, where some of the local product states are eigenstates of $X_l$ with $-1$ eigenvalue, the above $Q_l$ will be modified accordingly: $Q_l =e^{-\beta \lambda_B\sum_p^{l \in \partial p}  B_p   }  -  \eta_l    X_l$, where $\eta_l=-1$ for those links with $-1$ eigenvalue of $X_l$.

\subsection{3D toric code}
\subsubsection{Choosing $\ket{m}$ as a product state in Z basis}
In this case, the METTS is $\ket{\phi_m} \sim e^{-\beta H/2}\ket{m} \sim \prod_{s} \left( 1+A_s \tanh( \beta \lambda_A /2)   \right)\ket{m}$. Choosing $\ket{m}$ as a product state with eigenvalue $1$ for $Z_l$, the parent local Hamiltonian is $\widetilde{H}=\sum_l Q_l$ with $
Q_l =e^{-\beta  \lambda_A\sum_s^{l \in s}  A_s   }  -Z_l$.

\subsubsection{Choosing $\ket{m}$ as a product state in X basis}
In this case, the METTS is $\ket{\phi_m}\sim e^{-\beta H/2}\ket{m} \sim \prod_{p} \left( 1+B_p \tanh( \beta \lambda_B /2)   \right)\ket{m}$. Choosing $\ket{m}$ as a product state with eigenvalue $1$ for $X_l$, the parent Hamiltonian is $\widetilde{H}=\sum_l Q_l$ with $
Q_l =e^{-\beta \lambda_B \sum_p^{l \in \partial p}  B_p   }  -X_l$.

\subsection{4D toric code}
Below the indices $l$, $f$, and $c$ denote a link (1-cell), a face (2-cell), a cube (3-cell) respectively.
\subsubsection{Choosing $\ket{m}$ as a product state in Z basis}
In this case, the METTS is $\ket{\phi_m} \sim e^{-\beta H/2}\ket{m} \sim \prod_{l} \left( 1+A_l \tanh( \beta \lambda_A /2)   \right)\ket{m}$. Choosing $\ket{m}$ as a product state with eigenvalue $1$ for $Z_c$, the parent local Hamiltonian is $\widetilde{H}=\sum_f Q_f$ with $
Q_f =e^{-\beta  \lambda_A\sum_l^{l \in  \partial f}  A_l   }  -Z_f$.

\subsubsection{Choosing $\ket{m}$ as a product state in X basis}
In this case, the METTS is $\ket{\phi_m}\sim e^{-\beta H/2}\ket{m} \sim \prod_{p} \left( 1+B_c \tanh( \beta \lambda_B /2)   \right)\ket{m}$. Choosing $\ket{m}$ as a product state with eigenvalue $1$ for $X_f$, the parent Hamiltonian is $\widetilde{H}=\sum_f Q_f$ with $
Q_f =e^{-\beta \lambda_B \sum_p^{f \in \partial c}  B_c   }  -X_f$.

\subsection{Finite-depth quantum channel for a topologically trivial Gibbs state  }
Here we show that a finite-temperature Gibbs state of toric code models with the METTS decomposition $\rho=\sum_m p_m \ket{\phi_m} \bra{\phi_m}$ can be connected to a trivial classical mixed state using a finite-depth quantum channel as long as all $\ket{\phi_m}$ are short-range entangled. Below we focus on toric code models, but the argument applies to all CSS code Hamiltonians as well. First, we add an ancillary qubit (ancilla) on the location of every qubit, and consider a purified state on the enlarged Hilbert space as 
\begin{equation}
	\ket{\psi}= \sum_m \sqrt{p_m} \ket{\phi_m}\ket{m} \sim   \sum_m   \left[e^{\beta \lambda_A/2 \sum_s A_s }\otimes \mathbb{I}  \right] \ket{m}\ket{m},
\end{equation}
where $\ket{m}$ is a product state in Z basis, the first/second ket refers to the state in the system of interest/ancilla. Note that $T=1/\beta>T_A$ where $T_A$ is the critical temperature of $Z= \tr e^{\beta \lambda_A\sum_s A_s }$ so that $\ket{\phi_m}$ is short-range entangled. Denote $\ket{m}$ as a tensor product over lattice site $i$: $\ket{m}= \otimes_i \ket{m_i}$, $\ket{\psi}$ can be written as
\begin{equation}
	\ket{ \psi }  \sim \left[e^{\beta \lambda_A/2 \sum_s A_s }\otimes \mathbb{I}  \right]  \otimes_i \left[ \sum_{m_i} \ket{m_i} \ket{m_i}  \right],
\end{equation}
where $\otimes_i\left[ \sum_{m_i} \ket{m_i} \ket{m_i}  \right]$ is a tensor product of maximally entangled state at each site. Similar to the argument in the main text, the above equation implies that for $T>T_A$, $\ket{\psi }$ can be connected to $\otimes_i\left[ \sum_{m_i} \ket{m_i} \ket{m_i}  \right]$ without encountering a critical point, i.e. they can be connected using a finite-depth unitary. Therefore, we have proved the given $\rho$ can be connected to a trivial classical mixed state $ \sum_m\ket{m} \bra{m}$ using a finite-depth quantum channel.

\section{2D Toric Code}
\subsection{General formalism for taking partial transpose}

\begin{figure}
	\centering
	\begin{subfigure}[b]{0.26\textwidth}
		\caption{}
		\includegraphics[width=\textwidth]{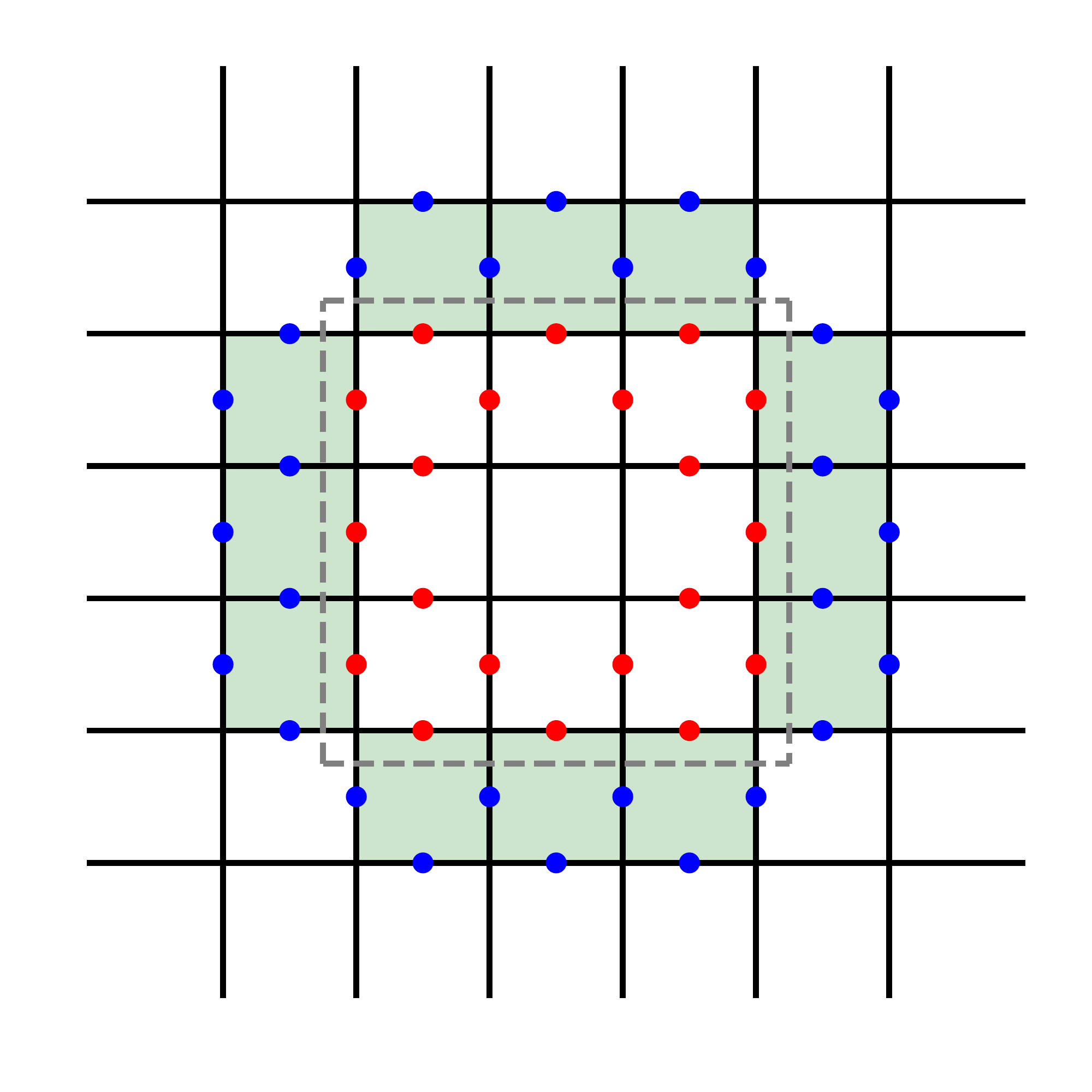}
		\label{fig:toric}
	\end{subfigure}
	\begin{subfigure}[b]{0.34\textwidth}
		\caption{}
		\includegraphics[width=\textwidth]{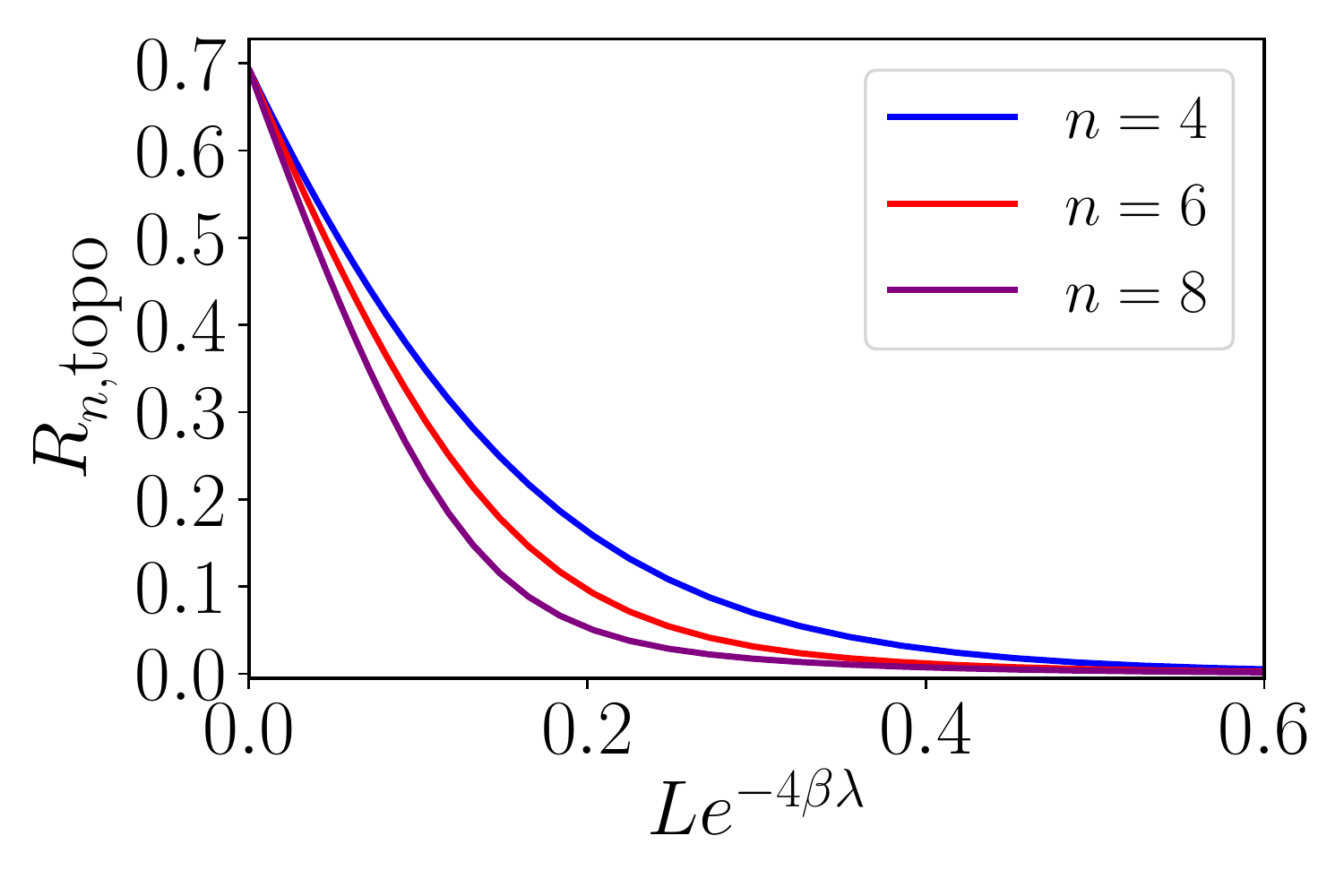}
		\label{fig:2drenyi_even}
	\end{subfigure}
	\begin{subfigure}[b]{0.34\textwidth}
		\caption{}
		\includegraphics[width=\textwidth]{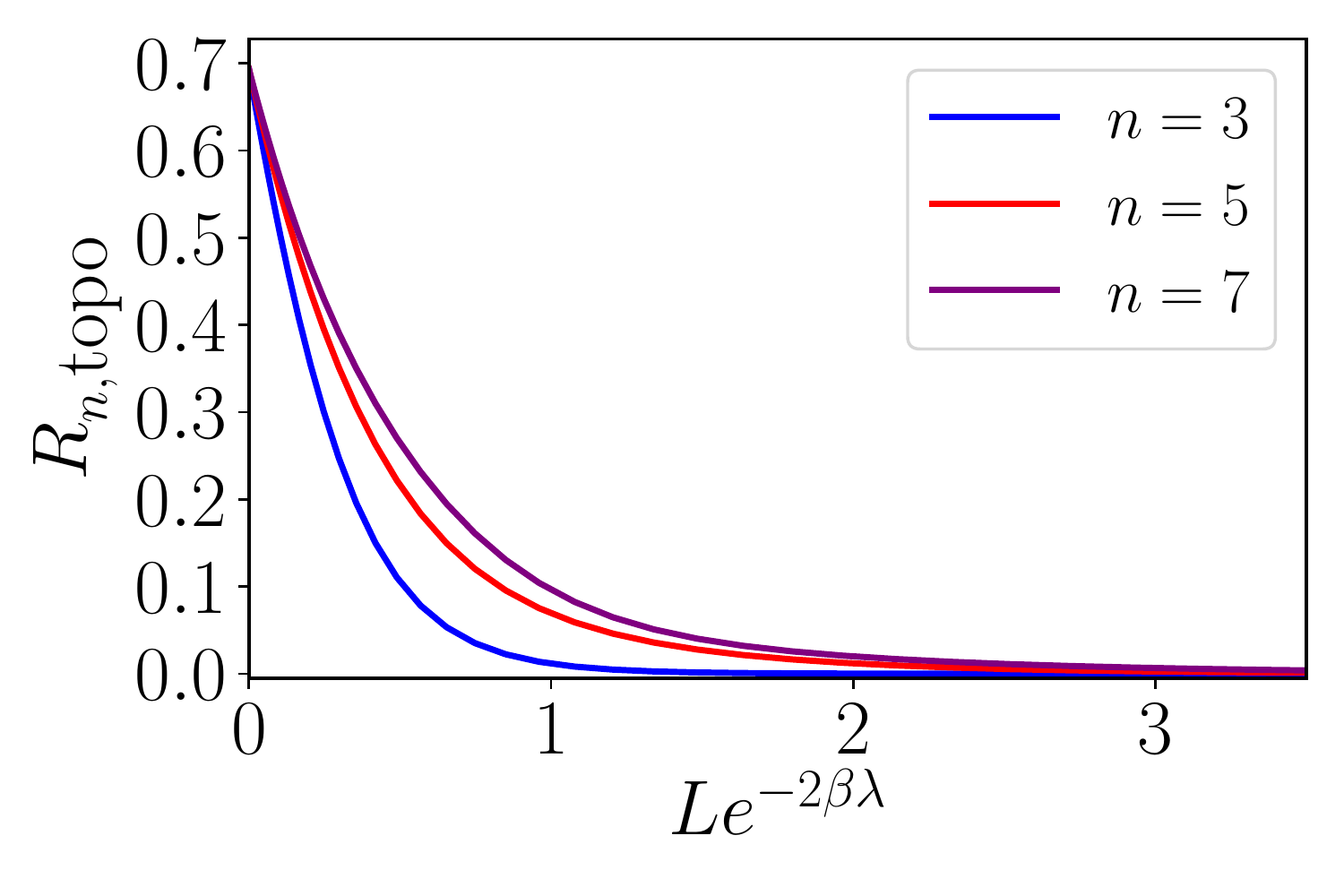}
		\label{fig:2drenyi_odd}
	\end{subfigure}
	\caption{ 2D toric code: (a)Every link has a spin-1/2 degree of freedom. The dashed gray line divides the system into $\mathcal{R}$ (inner region) and $\overline{\mathcal{R}}$ (outer region). The red and blue dots label the boundary spins involved in the boundary interaction $H_{\mathcal{R}\overline{\mathcal{R}}}$. (b)(c)The exact result of topological Renyi negativity for even $n$ and odd $n$ at $\lambda=\lambda_A=\lambda_B$ obtained using the generalized transfer matrix method without needing Monte Carlo sampling.}  
\end{figure}

A two dimensional toric code Hamiltonian reads
\begin{equation}
	H_T=-\lambda_A\sum_s A_s-\lambda_B \sum_{p} B_p,
\end{equation}
with $A_s$ and $B_p$ labelling the star and plaquette operators respectively. Dividing the lattice  into the region $\mathcal{R}$ and $\overline{\mathcal{R}}$ (Fig.\ref{fig:toric}), we write the toric code Hamiltonian as $H_T=H_\mathcal{R}+H_{\overline{\mathcal{R}}}+H_{R\overline{\mathcal{R}}}$, where $H_\mathcal{R}(H_{\overline{\mathcal{R}}})$ is supported only in $\mathcal{R}(\overline{\mathcal{R}})$ , and $H_{\mathcal{R} \overline{\mathcal{R}}}$ denotes the interaction between $\mathcal{R}$ and $\overline{\mathcal{R}}$. A thermal density matrix on the entire system is $\rho=\frac{1}{Z} e^{  -\beta (H_\mathcal{R}+H_{\overline{\mathcal{R}}}+H_{R\overline{\mathcal{R}}}  )  }$
with the thermal partition function $Z$:
\begin{equation}\label{eq:Z}
	Z=   \left( \cosh[N](\beta \lambda_A)+ \sinh[N](\beta \lambda_A)  \right)     \left( \cosh[N](\beta \lambda_B)+ \sinh[N](\beta \lambda_B)  \right)    2^{2N}.
\end{equation}
Here $N$ denotes the number of plaquette (or star) operators, and there are $2N$ spins in the system. Since partial transpose on $\rho$ only acts on the boundary part of the density matrix for commuting Hamiltonians, one finds $\rho^{T_{\overline{\mathcal{R}}}} =\frac{1}{Z} \left( e^{-\beta H_{\mathcal{R}\overline{\mathcal{R}}}}  \right)^{T_{\partial \overline{\mathcal{R}}}}  e^{-\beta  (H_\mathcal{R}+H_{\overline{\mathcal{R}}} )}$, where $\partial \overline{\mathcal{R}}$ denote the region in $\overline{\mathcal{R}}$ along the bipartition boundary. For the boundary interaction $H_{\mathcal{R}  \overline{\mathcal{R}}}$, there are $L$ boundary plaquettes and $L$ boundary stars, which can be labelled by index $i=1,2,\cdots, L$ clockwise. Therefore,

\begin{equation}\label{eq:part1}
	\begin{split}
		&  \left( e^{-\beta H_{ \mathcal{R}\overline{\mathcal{R}}  }} \right)^{T_{\partial \overline{\mathcal{R}}}}  \\
		&  =\left(  e^{ \beta \lambda_A\sum_{i=1}^L A_i +\beta \lambda_B \sum_{i=1}^L  B_i}   \right)^{T_{\partial \overline{\mathcal{R}}}}  \\
		& =\left[ \cosh \beta \lambda_A \right]^L  \left[\cosh\beta \lambda_B\right]^L  \sum_{\vec{n}} \sum_{\boldsymbol{\sigma}} \left( A_1^{n_1}\cdots A_L^{n_L}    B_1^{\sigma_1}\cdots B_L^{\sigma_L}   \right)^{T_{\partial \overline{\mathcal{R}}}} \left(   \tanh(\beta \lambda_A)\right)^{\sum_{i=1}^L n_s  }  \left(   \tanh(\beta \lambda_B)\right)^{\sum_{i=1}^L \sigma_s  },
	\end{split}
\end{equation}
where $\boldsymbol{n} = \{n_i \vert i=1, \cdots , L \} $ with $n_i=0,1$, and $\boldsymbol{\sigma} = \{\sigma_i \vert i=1, \cdots , L \} $ with $\sigma_i=0,1$. By introducing Ising variables $\tau_i=1-2n_i\in\{\pm 1\}$, indicating the absence or presence of star operators, one finds 
\begin{equation}
	\boxed{  \left(   A_1^{n_1}\cdots A_L^{n_L}    B_1^{\sigma_1}\cdots B_L^{\sigma_L}   \right)^{T_{\partial \overline{\mathcal{R}}}} =  A_1^{n_1}\cdots A_L^{n_L}  \left(  \tau_1\tau_2 B_1\right)^{\sigma_1} \cdots \left(  \tau_L\tau_1 B_L\right)^{\sigma_L}}.
\end{equation}
Plugging the above equation into Eq.\ref{eq:part1} and performing the summation over $\boldsymbol{\sigma}$ gives

\begin{equation}\label{eq:sub_partial}
	\left(  e^{-\beta H_{\mathcal{R} \overline{\mathcal{R}}}}   \right)^{T_{\partial \overline{\mathcal{R}}   }} = \cosh^L(\beta \lambda_A)   \sum_{   \boldsymbol{\tau}    }  e^{ -H  ( \boldsymbol{\tau} , \{A_i\} ,\{B_i\}   )},
\end{equation}
where $-H  (  \boldsymbol{\tau} ,\{A_i\} ,\{B_i\}   ) =  \sum_{i=1}^L     \frac{1-\tau_i}{2} \left(  \log A_{i} -K_A   \right)  +  \beta \lambda_B  B_i\tau_i\tau_{i+1}  $ and $K_A\equiv -\log \tanh(\beta \lambda_A)$. Therefore, the partial transposed density matrix is 
\begin{equation}\label{eq:appendix_partial_transpose}
	\rho^{T_{\overline{\mathcal{R}}}}= \frac{1}{Z}\left[ \cosh \beta \lambda_A \right]^L  e^{-\beta(H_\mathcal{R}+H_{\overline{\mathcal{R}}})} \sum_{   \boldsymbol{\tau}    } e^{ -H( \boldsymbol{\tau} , \{A_i\} ,\{B_i\}   )} .
\end{equation}
Note that $\left(  e^{-\beta H_{\mathcal{R} \overline{\mathcal{R}}}}   \right)^{T_{\partial \overline{\mathcal{R}}}}$ is essentially a partition function of 1D Ising model for $\tau_i$ spins, which can be calculated using the transfer matrix method: 

\begin{equation}
	\left(  e^{-\beta H_{\mathcal{R} \overline{\mathcal{R}}}}   \right)^{T_{\partial \overline{\mathcal{R}}}} 
	=\cosh^L(\beta\lambda_A)  \tr\left[ \prod_{i=1}^{L} M^{(i)}  \right]
\end{equation}
with the transfer matrix $M^{(i)}(A_i,B_i)$ is defined as 
\begin{equation}\label{append:transfer}
	M^{(i)}(A_i, B_i) = \begin{pmatrix}
		A_ie^{\beta \lambda_B B_i-K_A}   & A_ie^{-\beta \lambda_BB_i-K_A} \\
		e^{-\beta \lambda_BB_i}  & e^{\beta \lambda_BB_i} \end{pmatrix}
\end{equation} 
Given this result, the eigenspectrum of $\rho^{T_{\overline{\mathcal{R}}}}$ can be calculated efficiently.

\subsection{Renyi Negativity and Negativity}
We define the $n$-th Renyi Negativity $R_n$ as 
\begin{equation}\label{eq:renyi}
	R_n= b_n  \log \left(   \frac{  \tr{  \left(\rho^{T_{\overline{\mathcal{R}}}} \right)^n   }  }{ \tr\rho^n  }      \right),
\end{equation}
where $b_n=\frac{1}{1-n}$ for odd $n$ and  $b_n=\frac{1}{2-n}$ for even $n$. The presence of $\tr \rho^n$ within the logarithm makes $R_n$ obey an area law. Meanwhile, the prefactor is motivated by the fact that when $\rho$ is a pure state, $\tr{  \left(\rho^{T_{\overline{\mathcal{R}}}} \right)^n   }=  \tr_{\mathcal{R}} \rho_\mathcal{R}^n, \tr_\mathcal{R}\rho_{\mathcal{R}}^{\frac{n}{2}} $ for odd $n$ and even $n$ respectively. Thus $R_n$ reduces to Renyi entanglement entropy: $R_n=S_n,~S_{n/2}$ for odd $n$ and even $n$.
Define 
\begin{equation}\label{eq:appendix_define_r}
	r_n\equiv     \frac{  \tr{  \left(\rho^{T_{\overline{\mathcal{R}}}} \right)^n   }  }{ \tr\rho^n  }  =\frac{   \tr \{   \left(   e^{-\beta H}\right)^{T_{\overline{\mathcal{R}}}}       \}^n   }{    \tr e^{-n\beta H}},
\end{equation}
using Eq.\ref{eq:appendix_partial_transpose}, we find 
\begin{equation}
	r_n=  \left[\cosh\beta \lambda_A \right]^{nL}  \frac{  \sum_{\{A_s\} ,\{B_p\}   }  f(\{A_s\},\{B_p\})   e^{n\beta \lambda_A \sum^{\text{bulk}}_{s} A_s +n\beta\lambda_B  \sum^{\text{bulk}}_{p}B_p }    \left( \sum_{\{\tau_i\}  }      e^{-H}\right)^n     }{      \sum_{\{A_s\} ,\{B_p\}   }  f(\{A_s\},\{B_p\})  e^{n\beta \lambda_A \sum_{s} A_s +n\beta\lambda_B  \sum_{p}B_p }   },
\end{equation}
where $ f(\{A_s\},\{B_p\})  $ gives the global constraint: $f(\{A_s\},\{B_p\}) =  \delta\left(\prod_{s}A_s =1\right)  \delta\left(\prod_{p}B_p =1\right)$. Summing $A_s$ and $B_p$ in the bulk gives
\begin{equation}
	r_n=   \left[\frac{  \left(     \cosh\beta \lambda_A \right)^n}{4} \right]^L \sum^{\partial }_{A_s} \sum^{\partial }_{B_p} W\left( \prod^{\partial}_s A_s , \prod^{\partial}_p B_p \right) \left(         \sum_{\{\tau_i\}} e^{-H}  \right)^n, 
\end{equation}
where 
\begin{equation}
	W\left( \prod^{\partial}_s A_s , \prod^{\partial}_p B_p \right)=\frac{ \left\{   \left[\cosh n\beta \lambda_A \right]^{N-L}  +   \left[\sinh n\beta \lambda_A \right]^{N-L} \prod^{\partial}_s A_s    \right\}   \left\{   \left[\cosh n\beta \lambda_B \right]^{N-L}   +  \left[\sinh n\beta \lambda_B \right]^{N-L}   \prod^{\partial}_p B_p  \right\}     }{    \left\{     \left[\cosh n\beta\lambda_A \right]^N +   \left[\sinh n\beta\lambda_A \right]^N \right\}      \left\{     \left[\cosh n\beta\lambda_B \right]^N +   \left[\sinh n\beta\lambda_B \right]^N \right\}              }. 
\end{equation}
As $N\to \infty $, $W$ reduces to $\left[    \cosh(n\beta \lambda_A) \cosh( n\beta \lambda_B) \right]^{-L}$, and thus we find

\begin{equation}\label{eq:appendix_r_n}
	r_n =  \frac{1}{ 2^{2L}} \left(  \frac{  \left[\cosh\beta\lambda_A \right]^n }{\cosh(n\beta \lambda_A) \cosh(n\beta \lambda_B)      }   \right)^L  \sum_{ \substack{  \{   A_i=\pm 1\\  i=1,\cdots, L   \} }  } \sum_{\substack{  \{B_i=\pm 1 \\  i=1,\cdots ,L \}} }        \left(  \sum_{ \{ \tau_i \}    }  e^{-H } \right)^n , 
\end{equation}
where $-H=  \sum_{i=1}^L     \frac{1-\tau_i}{2} \left(  \log A_{i} -K_A   \right)  +  \beta \lambda_B  B_i\tau_i\tau_{i+1}  $ and $K_A\equiv -\log \tanh(\beta \lambda_A)$.

\subsubsection{Negativity}
The analytic continuation gives the negativity $E_N= \lim_{     \text{even }n \to 1   }R_n$:
\begin{equation}\label{eq:nega}
	E_N=\log \expval{   \abs{ Z(\{ A_i  \},   \{B_i\}) }},
\end{equation}
where $Z(\{A_i  \},   \{B_i\})=\frac{1}{ \left[ \cosh\beta \lambda_B\right]^L} \sum_{ \{ \tau_i=\pm 1\}} e^{-H   ( \boldsymbol{\tau}  , \{  A_i \} , \{B_i  \}   )    }$ with the angled brackets denoting the disorder average over the variables $\{A_i\}$ and $\{B_i\}$. Since $Z=\frac{1}{  \left[  \cosh(\beta \lambda_B)   \right]^L} \tr\left[\prod_{i=1}^L M^{(i)} \right]$, where $M^{(i)}$ is the transfer matrix defined in Eq.\ref{append:transfer}, we can efficiently calculate $Z$ for a given $\{A_i\}$ and $\{B_i  \}$, and perform Monte Carlo sampling over various $\{A_i\}$ and $\{B_i  \}$ to calculate $E_N$ (see Fig.2 inset in the main text for result).

\subsubsection{Generalized transfer matrix method}
Renyi negativity $R_n$ can be calculated exactly without needing Monte Carlo sampling. The trick is to introduce a generalized transfer matrix $\widetilde{M}$ of dimension $2^n\cross 2^n$: $\widetilde{M}=\sum_{A=\pm 1,B= \pm 1}  M^{\otimes n } (A,B)$, where $M$ is defined in Eq.\ref{append:transfer}. It follows that

\begin{equation}
	r_n=\frac{  \tr{  \left(\rho^{T_{\overline{\mathcal{R}}}} \right)^n   }  }{ \tr\rho^n  }=  \left(  \frac{  \cosh^n(\beta\lambda_A)  }{4\cosh(n\beta \lambda_A) \cosh(n\beta \lambda_B)      }   \right)^L    \tr \widetilde{M}^L   ,
\end{equation}
As a result, Renyi negativity $R_n$ can be calculated for arbitrary $\beta,\lambda_A, \lambda_B$ by finding the eigenvalues $\lambda_i$ of the generalized transfer matrix $\widetilde{M}$. Define the sorted eigenvalues $\lambda_1\geq \lambda_2\geq  \cdots \geq \lambda_{2^n}$, Renyi negativity can be written as $R_n= \alpha_n L- R_{n,\text{topo}}$ with the area-law coefficient $\alpha_n=b_n  \log  \left(\frac{ \lambda_1 \cosh[n](\beta\lambda_A)  }{4\cosh(n\beta \lambda_A) \cosh(n\beta \lambda_B)      }  \right)$ and the topological Renyi negativity $R_{n,\text{topo}}=-b_n \log \left[  1+\sum_{i=2}^{2^n}    \left(  \frac{\lambda_i}{\lambda_1}  \right)^L  \right]$ (See  Fig.\ref{fig:2drenyi_even},\ref{fig:2drenyi_odd} for result).

\subsubsection{Low temperature expansion for general $\lambda_A$ and $\lambda_B$}
In $\beta \to \infty$ limit for finite $L$, the subleading term $\gamma_n$ follows the expression:
\begin{equation}
	R_{n,\text{topo}}=\begin{cases}
		\log 2 -\frac{L}{2} \left(  e^{-2\beta\lambda_A} +e^{-2\beta \lambda_B} \right) - L\sqrt{e^{-4\beta \lambda_A}   -    e^{-2\beta( \lambda_A+\lambda_B  )}  + e^{-4\beta \lambda_B} } +\cdots \quad ~\text{for } \quad n=3   \\
		\log 2 - L\left(  e^{-4\beta\lambda_A} +e^{-4\beta \lambda_B} \right) - 2L\sqrt{e^{-8\beta \lambda_A}   -    e^{-4\beta( \lambda_A+\lambda_B  )}  + e^{-8\beta \lambda_B} } +\cdots \quad \text{for }  \quad n=4
	\end{cases}
\end{equation}
One can set $\lambda=\lambda_A=\lambda_B$ and find that $R_{n,\text{topo}}$ depends on $Le^{-4\beta\lambda}$ and $Le^{-2\beta\lambda}$ for even $n$ and odd $n$ respectively (consistent with Fig.\ref{fig:2drenyi_even},\ref{fig:2drenyi_odd}).

\subsubsection{$\lambda_B\to \infty$ limit}
In this limit, we can obtain a compact expression for negativity and Renyi negativity. To begin with ,
\begin{equation}
	\sum_{ \substack{  \{   A_i=\pm 1\\  i=1,\cdots, L   \} }  } \sum_{\substack{  \{B_i=\pm 1 \\  i=1,\cdots ,L \}} }        \left(   \sum_{\substack{  \{\tau_i=\pm 1 \\  i=1,\cdots ,L \}} }    e^{-H}   \right)^n  = 
	\sum_{ \substack{  \{   A_i=\pm 1\\  i=1,\cdots, L   \} }  } \sum_{\substack{  \{B_i=\pm 1 \\  i=1,\cdots ,L \}} }     \left( \sum_{\substack{  \{\tau_i=\pm 1 \\  i=1,\cdots ,L \}} }   e^{\sum_{s=1}^L    \beta \lambda_B B_i\tau_i\tau_{i+1}  -\frac{1}{2} (K_A- \log A_i) (1-\tau_i)      }   \right)^n . 
\end{equation}
As $\lambda_B \to \infty$, we only needs to consider the $\tau_i$ spin configurations satisfying $B_i  \tau_i\tau_{i+1}=1$, and thus for a given $\{B_i   \}$, only two $\{\tau_i\}$ configurations related by global spin flips are allowed. Also note that $\prod_{i=1}^L B_i$ needs to be one so that there is no frustration for $\{\tau_i\}$. A straightforward calculation shows that

\begin{equation}\label{eq:append_r_n}
	\begin{split}
		r_n  &=  \left(   \frac{\cosh[n](\beta \lambda_A)}{\cosh(n\beta \lambda_A)}  \right)^L    \sum_{m=0,2,4,\cdots}^{n}   \binom{n}{m}  e^{-K_A m } \left[ e^{-K_Am }  +e^{-K_A (n-m)} \delta( \text{even}~n   )   \right]^{L}\\
		&= \begin{cases}
			\left(   \frac{\cosh[n](\beta \lambda_A)}{\cosh(n\beta \lambda_A)}  \right)^L  \sum_{m=0,2,4,\cdots}^{n}    \binom{n}{m}  e^{-LK_A m }\quad  \text{for odd } \quad n    \\
			\frac{1}{2}  \left(   \frac{\cosh[n](\beta \lambda_A)}{\cosh(n\beta \lambda_A)}  \right)^L  \sum_{m=0,2,4,\cdots}^{n}    \binom{n}{m} \left[ e^{-K_Am }  +e^{-K_A (n-m)}   \right]^{L} \quad    \text{for  even } \quad n.
		\end{cases}
	\end{split}
\end{equation}
\textbf{\textit{Odd n}}
\\
For odd $n$, from Eq.\ref{eq:append_r_n}, one finds Renyi negativity $
R_n=\frac{L}{1-n}  \log \left[    \frac{\cosh[n](\beta \lambda_A)}{\cosh(n\beta \lambda_A)}     \right]   -R_{n,\text{topo}}$, where 
\begin{equation}
	\boxed{     R_{n,\text{topo}}  =  \frac{1}{n-1}  \log \left\{  \frac{1}{2} \left[   \left(    1+\tanh[L](\beta \lambda_A) \right)^n     +     \left(   1-\tanh[L](\beta \lambda_A) \right)^n     \right]   \right\}}  .
\end{equation}
At large $\beta \lambda_A$, one finds $R_{n,\text{topo}}=\log 2 -\frac{n}{n-1}Le^{-2\beta \lambda_A} +\cdots$.

\vspace{5mm}
\noindent\textbf{\textit{Even n}}\\
From Eq.\ref{eq:append_r_n}, one finds
\begin{equation}
	R_n= \frac{L}{2-n} \log \left\{\frac{   \left(1+ \tanh[n](\beta \lambda_A)\right) \cosh[n](\beta \lambda_A)   }{ \cosh(n\beta \lambda_A)}    \right\} - R_{n,\text{topo}},
\end{equation}
where the subleading term $R_{n,\text{topo}}$ is
\begin{equation}\label{eq:appendix_gamma_n_ series}
	\boxed{   R_{n,\text{topo}}= - \frac{1}{2-n} \log \left\{    \frac{1}{2} \sum_{m=0,2,4,\cdots}^{n}   \binom{n}{m}\left(      \frac{     \tanh[m](\beta \lambda_A)   + \tanh[n-m](\beta \lambda_A) }{ 1+\tanh[n](\beta \lambda_A)    }     \right)^L  \right\}  }   .
\end{equation}

\noindent For $n>2$, $R_{n,\text{topo}}$ can be written as 

\begin{equation}
	R_{n,\text{topo}}=- \frac{1}{2-n} \log \left\{ 1+   \frac{1}{2} \sum_{m=2,4,\cdots}^{n-2}   \binom{n}{m}\left(      \frac{     \tanh[m](\beta \lambda_A)   + \tanh[n-m](\beta \lambda_A) }{ 1+\tanh[n](\beta \lambda_A)    }     \right)^L  \right\} .
\end{equation}
As one can check, taking $\beta\to \infty $ first before sending $L\to \infty$ gives $R_{n,\text{topo}}=\log 2$ while sending $L\to \infty $ first for any nonzero temperature gives $R_{n,\text{topo}}=0$. At large $\beta \lambda_A$ limit, this expression can be simplified as $R_{n,\text{topo}}=  \log 2 -  \frac{n(n-1)}{2(n-2)} Le^{-4\beta \lambda_A}+\cdots$.

\vspace{5mm}
\noindent\textbf{\textit{Analytic continuation of $R_{n,\text{topo}}$ of even $n$ as $n\to 1$}}\\
Starting with Eq.\ref{eq:appendix_gamma_n_ series}, we can take $n\to 1$ to obtain the subleading term in the negativity, i.e. topological entanglement negativity. To begin with, we expand  $\left(  \tanh^m(\beta \lambda_A)   + \tanh^{n-m}(\beta \lambda_A) \right)^L$, and write 
\begin{equation}
	\sum_{m=0,2,4,\cdots}^{n}   \binom{n}{m} \left(    \tanh[m](\beta \lambda_A)   + \tanh[n-m](\beta \lambda_A)    \right)^L=\sum_{m=0}^{n}\frac{\left(  1+(-1)^m  \right)}{2} \sum_{k=0}^L \binom{L}{k} \left[\tanh(\beta \lambda_A)\right]^{m(2k-L)+n(L-k)   }.
\end{equation}
Performing the summation over $m$, we find $R_{n,\text{topo}}$ is equal to 
\begin{equation}
	\frac{1}{n-2}  \log  \left\{   \frac{1}{4}  \left(  1+\tanh^n(\beta \lambda_A)  \right)^{-L}   \sum_{k=0}^L \binom{L}{k}  \left[    \left( \tanh^{L-k}(\beta\lambda_A)  +  \tanh^{k}(\beta\lambda_A)        \right)^n +   \left( \tanh^{L-k}(\beta\lambda_A)  -  \tanh^{k}(\beta\lambda_A)        \right)^n  \right]           \right\}.
\end{equation}
By analytically continuing even $n\to 1$, one finds 

\begin{equation}
	E_{N,\text{topo}} =  - \log  \left\{  \frac{1}{4}     \left(  1+\tanh(\beta \lambda_A)  \right)^{-L}       \sum_{k=0}^L \binom{L}{k} \left(  \tanh^{L-k}(\beta\lambda_A)  +  \tanh^{k}(\beta\lambda_A)   +  \abs{  \tanh^{L-k}(\beta\lambda_A)  -  \tanh^{k}(\beta\lambda_A)    }    \right)    \right\}.
\end{equation} 
To proceed, we define $x=\tanh(\beta\lambda_A)$ for notational convenience. Removing the absolute value sign results in a truncated binomial summation 
\begin{equation}
	E_{N,\text{topo}} = - \log  \left\{  \frac{1}{2}       +  \frac{1}{2}\left(  1+x  \right)^{-L}    \sum_{k=0}^{L/2}  \binom{L}{k} (    x^k  -x^{L-k} )                 \right\}.
\end{equation}
As $\beta\to1$, $x\to 1$, $E_{N,\text{topo}} $ reduces to $\log 2$, consistent with the topological entanglement entropy in the ground state. As $L \to \infty$, the truncated binomial summation can be performed exactly via a saddle point calculation, showing $E_{N,\text{topo}} = 0$. In fact, the truncated binomial summation can be calculated exactly by introducing the hypergeometric function $_2F_1$:
\begin{equation}
	\sum_{k=0}^{L/2}  \binom{L}{k}  x^k=\left(  1+x  \right)^L -x^{1+L/2} \binom{L}{\frac{L}{2}+1}  ~ _2F_1( 1,-\frac{L}{2}+1 ;  \frac{L}{2}+2 ;-x ),
\end{equation}
and thus $E_{N,\text{topo}} $ allows a compact expression:
\begin{equation}
	\boxed{  E_{N,\text{topo}} = - \log   \left\{     \frac{1}{2}    +   \frac{1}{2} \left(  x^{1/2}+x^{-1/2} \right)^{-L}    \binom{L}{\frac{L}{2}+1}    \left[ \frac{1}{x} ~  _2F_1( 1,-\frac{L}{2}+1 ;  \frac{L}{2}+2 ;-\frac{1}{x}  )   -  x  ~ _2F_1( 1,-\frac{L}{2}+1 ;  \frac{L}{2}+2 ;-x )       \right]     \right\} }
\end{equation}
For finite $L$, a low temperature expansion gives 
\begin{equation}
	E_{N,\text{topo}} =  \log 2-\frac{L+2}{2^L} \binom{L}{\frac{L}{2}+1} e^{-2\beta \lambda_A}  +\frac{2}{\pi} \left(  \frac{ \Gamma\left(\frac{L+1}{2}   \right)   }{ \Gamma\left(\frac{L}{2}  \right)  }  \right)^2 e^{-4\beta \lambda_A}+ O(e^{-6\beta \lambda_A}).
\end{equation}
For large $L\gg 1$, one can employ Stirling's approximation for further simplification:
\begin{equation}
	E_{N,\text{topo}} = \log 2 - \sqrt{   \frac{2}{\pi}   Le^{-4\beta \lambda_A}  }   +\frac{L}{\pi} e^{-4\beta \lambda_A}  + O(e^{-6\beta \lambda_A}),
\end{equation}
which depends on the scaling variable $Le^{-4\beta \lambda_A}$, consistent with Fig.2 in the main text.

\section{3D Toric Code}

\subsection{General expression of (Renyi) negativity}
In a three dimensional lattice of size $L_x\cross L_y \cross L_z$ we consider a three dimensional toric code Hamiltonian 
\begin{equation}
	H_T= -\lambda_A \sum_{s}A_s -\lambda_B \sum_p B_p,
\end{equation}
with $A_s=\prod_{i\in s} X_i$ and $B_p=\prod_{i\in p} Z_p$. This model has the local constraint on each cubic unit cell: $\prod_{p\in\text{cube}} B_p=1$. There can be global constraint for $A_s$ and $B_p$ as well, depending on the topology of the lattice. For simplicity, we impose periodic boundary condition along $x$ and $y$ direction, while impose open boundary condition along $z$ direction. This boundary condition gives the extra global constraint: the product of $B_p$ on $x-y$ plane gives identity operator. The modification of our subsequent result will be straightforward when considering different boundary conditions.

Our first goal is to find  partial transposed density matrix: $ \rho^{T_{\overline{\mathcal{R}}}}  =\frac{1}{Z}   \left( e^{-\beta H_{\mathcal{R}\overline{\mathcal{R}}} } \right)^{T_{\partial \overline{\mathcal{R}}}}  e^{-\beta \left(  H_\mathcal{R}+H_{\overline{\mathcal{R}}} \right)  } $. We choose the subsystem $\mathcal{R}$ separated from its complement $\overline{\mathcal{R}}$ by a two dimensional plane at $z=z_0$. Let's first study $   \left( e^{-\beta H_{\mathcal{R}\overline{\mathcal{R}}} } \right)^{T_{\partial \overline{\mathcal{R}}}}$, we will show that its is equivalent to the partition function of a two dimensional random-bond Ising model with random on-site complex magnetic field. Labelling a lattice site by $i$ on the two dimensional boundary plane, there are $N_s^{\partial}=L_xL_y$ number of boundary star operator $A_i$, and $N_p^{\partial}= 2L_xL_y$ number of boundary plaquette operators $B_{ij}$ living on the link $\expval{ij}$. Using the equality 

\begin{equation} 
	\left\{  \left[  \prod_{i}    A_{i}^{n_i}   \right]    \left[  \prod_{ \expval{ij}  }   B_{ij}^{\sigma_{ij}}   \right] \right\}^{T_{\partial \overline{\mathcal{R}}}}   =   \left[  \prod_{i} A_i^{n_{i}}   \right]    \left[  \prod_{\expval{ij}}  \left(  \tau_{i}\tau_{j}  B_{ij}   \right)^{\sigma_{ij}}       \right],
\end{equation}
analogous to 2D toric code with $\tau_{i}=1-2n_{i}$, we find $
\left( e^{-\beta H_{\mathcal{R}\overline{\mathcal{R}}} } \right)^{T_{\partial \overline{\mathcal{R}}}}= \cosh[N_s^{\partial }](\beta \lambda_A)   \sum_{\{ \tau_{i} \}} e^{-H}$, where
\begin{equation}\label{append:expression1_H}
	-H( \{A_i   \}  ,  \{ B_{ij}  \} , \{ \tau_{i} \})  = \sum_{ i} \frac{1-\tau_{i}}{2}  \left(\log A_{i} -K_A \right) + \beta \lambda_B \sum_{\expval{ij}}    B_{ij}  \tau_{i} \tau_{j} 
\end{equation}
with $K_A\equiv -\log  (\tanh(\beta \lambda_A))$. With the above equation, we derive the moment for Renyi negativity:
\begin{equation}\label{append:3d_general}
	\begin{split}
		r_n&=\frac{\tr \left\{  e^{-n\beta (H_\mathcal{R}+H_{\overline{\mathcal{R}}})}  \left[  \left(e^{-\beta H_{\mathcal{R}\overline{\mathcal{R}}}}\right) ^{T_{\partial \overline{\mathcal{R}}}}\right]^n  \right\} }{  \tr   \left[ e^{-n\beta \left(   H_\mathcal{R}+H_{\overline{\mathcal{R}}}+H_{\mathcal{R}\overline{\mathcal{R}}} \right)}   \right]  }\\
		&= \frac{   \sum_{\{A_s\}} \sum_{\{ B_p \}} f( \{A_s\}, \{B_p  \}    )  e^{    -n\beta(H_\mathcal{R}+H_{\overline{\mathcal{R}}}) }  \cosh^{nN_s^{\partial }}(\beta \lambda_A)     \left[      \sum_{\{ \tau_i \}} e^{-H} \right]^n   }{       \sum_{\{A_s\}} \sum_{\{ B_p \}}  f( \{A_s\}, \{B_p  \}    ) e^{    -n\beta(H_\mathcal{R}+H_{\overline{\mathcal{R}}}+H_{\mathcal{R}\overline{\mathcal{R}}}) } },
	\end{split}
\end{equation}
where $f( \{A_s\}, \{B_p  \}    )$ imposes the constraint

\begin{equation}\label{append:global_constraint}
	f( \{A_s\}, \{B_p  \}    )=  \delta\left( \prod_{p \in x-y ~\text{plane}} B_p=1    \right) \prod_{\text{cube}} \delta\left(   \prod_{p\in \text{cube} } B_p=1 \right).
\end{equation}
Suppose one considers a three-torus, the extra constraint $\delta\left(  \prod_s A_s =1  \right) \delta\left( \prod_{p \in y-z ~\text{plane}} B_p=1    \right) \delta\left( \prod_{p \in z-x~ \text{plane}}  B_p=1   \right) $ will also be present in Eq.\ref{append:global_constraint}. Furthermore, one will need to boundaries $z=z_1$, $z=z_2$ to divide the three-torus, but the corresponding modification for Eq.\ref{append:3d_general} is straightforward.

\subsection{$\lambda_B\to \infty $ limit}
In this limit, for the denominator of $r_n$, all the plaquette operators $B_p=1$. For the numerator, all the plaquette operators $B_p$ in the bulk is pinned at one, while $B_p$ across the boundary and $\tau_{i}$ degrees of freedom satisfy $B_{ij} \tau_{i}\tau_{j}=1$. Furthermore, to make the above conditions hold true for all bonds, i.e. no frustration for $\tau_{i}$ spins, the following  constraint $f_{\partial }$ for boundary $B_p$ is required:
\begin{equation}
	f_{\partial }( \{  B_i\} ) =  \delta\left( \prod_{\expval{ij} \in \Gamma_x  } B_{ij} =1 \right)  \delta\left( \prod_{\expval{ij} \in \Gamma_y  } B_{ij} =1 \right)  \prod_{\square} \delta\left(   \prod_{\expval{ij}\in \square}  B_{ij}=1 \right),
\end{equation}
where $\Gamma_{x(y)}$ denotes a non-contractible loop along $x(y)$ direction. Note that they arise due to the periodic boundary condition in $x$ and $y$ direction. On the other hand, $\prod_{\expval{ij} \in \square}$ denotes a product of four links surrounding a plaquette unit cell. Therefore, 
\begin{equation}
	r_n=\frac{1}{2^{N_s^{\partial }}}  \left[  \frac{\left(   \cosh\beta \lambda_A\right)^n}{\cosh n\beta \lambda_A}\right]^{N_{s}^{\partial}}\sum^{}_{\{A_i\}} \sum^{}_{\{B_{ij} \}}  f_{\partial }( \{  B_i\} )  \left[    e^{\sum_{i}        \frac{1}{2}(1+\tau_{i}   ) \left( \log A_{i} -K_A \right)      }  +e^{\sum_{i}        \frac{1}{2}(1-\tau_{i}   ) \left( \log A_{i} -K_A \right)      } \right]^n, 
\end{equation}
where one of the spin, say $\tau_1$ is fixed at one, and the rest  $\tau_{i}$ are determined by $B_{ij}$ accordingly from the constraint $B_{ij} \tau_{i}\tau_{j}=1$. This implies that there are $2N_{s}^{\partial}-(N_{s}^{\partial} +1 )= N_{s}^{\partial}-1$ independent boundary plaquette operators. By converting the summation over those independent $B_{ij}$ to the summation over $\tau_{i}$ spins (except for $\tau_{1}$), and a little bit calculation, we find in $\lambda_B\to \infty$ limit, (Renyi) negativity in the 3D toric code is simply given by that in the 2D toric code by replacing boundary length $L$ in the 2D toric code by the boundary area $N_s^{\partial }=L_xL_y$ in the 3D toric code.

\subsection{$\lambda_A\to \infty $ limit}
In this limit, we will perform an low temperature perturbative calculation for Renyi negativity. To avoid ambiguous excitations arising from the boundary of the cubic lattice under open boundary condition, here we consider three-torus topology. This requires two planes $z=z_1$, $z=z_2$ to divide the system into two parts. We find that the two disconnected boundary planes contribute independently (and equally) to negativity. Thus we will just focus on the negativity contribution from one of the planes.

To begin with, $\lambda_A\to \infty $ makes $A_s=1$ in the denominator of $r_n$, while in the numerator, only the bulk $A_s=1$. Thus 

\begin{equation}
	r_n= 2^{-nA}    \frac{ \sum_{\{ A_i  \}} \sum^{\text{bulk}}_{\{B_p \}}  \sum_{  \{ B_{ij} \} } f(\{ B_p\})  e^{n\beta \lambda_B   \sum^{\text{bulk}}_p  B_p}      \left(   \sum_{\{  \tau_i  \}}  \prod_{i}A_{i}^{\frac{1-\tau_{i}}{2}}   e^{\beta \lambda_B  \sum_{\expval{ij}} B_{ij} \tau_{i}\tau_j   }   \right)^n        }{  \sum_{\{B_p\}}  f(\{ B_p\}) e^{ \sum_{p}n\beta \lambda_B  B_p}          }.
\end{equation}
Summing over $\{A_i\}$ in the numerator gives
\begin{equation}
	\sum_{\{ A_i\}}    \left(   \sum_{\{  \tau_i  \}}  \prod_{i}A_{i}^{\frac{1-\tau_{i}}{2}}   e^{\beta \lambda_B  \sum_{\expval{ij}} B_{ij} \tau_{i}\tau_j   }   \right)^n 
	=2^A \sum_{\{\tau^{1}_i \}  }\cdots  \sum_{\{\tau^{n}_i \}  } \prod_i \delta\left(  \prod_{\alpha=1}^n\tau_i^{\alpha} =1   \right)    e^{\beta \lambda_B \sum_{\alpha=1}^n   \sum_{\expval{ij}} B_{ij}  \tau^{\alpha}_{i}\tau_j^{\alpha}   }.
\end{equation}
This is essentially a partition function for $n$ replicas of the two dimensional Ising model, where the spins in different replicas at any given lattice site index $i$ are coupled through the delta function constraint. Therefore, $r_n =2^{A(1-n)}  \tilde{Z}/Z$, where 
\begin{equation}
	\begin{split}
		&\tilde{Z}=   \sum^{\text{bulk}}_{\{B_p \}}  \sum_{  \{ B_{ij} \} }  \sum_{\{\tau^{1}_i \}  }\cdots  \sum_{\{\tau^{n}_i \}  }  f(\{ B_p\})  e^{n\beta \lambda_B   \sum^{\text{bulk}}_p  B_p}   \prod_i \delta\left(  \prod_{\alpha=1}^n\tau_i^{\alpha} =1   \right)    e^{\beta \lambda_B \sum_{\alpha=1}^n   \sum_{\expval{ij}} B_{ij}  \tau^{\alpha}_{i}\tau_j^{\alpha}   }  \\
		&Z=  \sum_{\{B_p\}}  f(\{ B_p\}) e^{n\beta \lambda_B   \sum_p B_p}.
	\end{split}
\end{equation}
Hence up to the prefactor $2^{A(1-n)}$, $r_n$ is essentially the ratio of two partition function $\tilde{Z}$ and $Z$,  where the partition function $\tilde{Z}$ describes a 3D $Z_2$ gauge theory of inverse temperature $n\beta $ coupled to n replicas of 2D Ising models, and $Z$ describes the conventional 3D $Z_2$ gauge theory. Using these equations, one can then perform a conventional low temperature expansion for $\tilde{Z}/Z$ to study $r_n$. Since $\tilde{Z}$ and $Z$ both have a finite temperature critical point given by 3D $Z_2$ gauge theory, the perturbative expansion is gauranteed to be convergent at low temperature.

To gain intuition, we first consider the zero temperature limit, where one only needs to consider the ground states. $Z$ has a unique ground state by setting $B_p=1~\forall p$, hence $Z=e^{n\beta 3L^3}$. For $\tilde{Z}$, maximizing the Boltzmann weight requires $B_p=1$ in the bulk, and $B_{ij} \tau^{\alpha}_i \tau^{\alpha}_j=1 ~ \forall \expval{ij}$. This implies 
\begin{equation}
	\tilde{Z}  =  e^{n\beta 3L^3} g,
\end{equation}
where $g$ is the ground state degeneracy given by
\begin{equation}  \sum_{  \{ B_{ij} \} }  \sum_{\{\tau^{1}_i \}  }\cdots  \sum_{\{\tau^{n}_i \}  } \prod_i \delta\left(  \prod_{\alpha=1}^n\tau_i^{\alpha} =1   \right)      \prod_{\alpha=1}^n\prod_{\expval{ij}} \delta\left( B_{ij}\tau^{\alpha}_i\tau^{\alpha}_j=1\right)
\end{equation}

Below we will consider Renyi negativity for even $n$ and odd $n$ separately.
\begin{itemize}
	\item even $n$	
\end{itemize}

Finding the number of ground states correspond to $\tilde{Z}$ is equivalent to finding the dimension of the symmetry group of the action of $\tilde{Z}$. First we find the global spin flip ($Z_2$) symmetry by sending $\tau_i^{\alpha} \to -\tau_i^{\alpha} ~\forall i$ on a given replica. However, due to the constraint between different replicas $\prod_i \delta\left( \prod_{\alpha=1}^n \tau_i^{\alpha} =1\right)$, only the symmetry generator for $n-1$ replicas can be freely chosen. Hence the dimension of this group is $2^{n-1}$. Apart from the global spin flip symmetry, the action of $\tilde{Z}$ also possesses local symmetry. A local symmetry generator $G_i$ takes $B_{ij}\to -B_{ij} $ for four $B_{ij}$ emanating from a given site $i$, and $\tau_i^{\alpha} \to -\tau^{\alpha}_{i} ~\forall \alpha$ at the same time. Notice this respects the constraint $\prod_i \delta\left( \prod_{\alpha=1}^n \tau_i^{\alpha} =1\right)$ since $n$ is even. The dimension of local symmetry operation is $2^{A-1}$ since $\prod_{i}G_i$ acts trivially on $B_{ij}$ ( it acts non-trivially on spins but the effect has been accounted by the aforementioned global spin flips). As a result, the total dimension of the symmetry group ( and the ground state degeneracy) for the action of $\tilde{Z}$ is $2^{A-1}2^{(n-1)}$. This implies

\begin{equation}
	r_n=2^{ A(1-n) }2^{A-1}2^{(n-1)} =  2^{(A-1)(2-n)},
\end{equation}
giving the ground state negativity $R_n =\frac{1}{2-n} \log r_n   =  A \log 2-\log 2$. Next by considering the excitations from ground states, we calculate $r_n$ using low temperature expansion. Note that this perturbative expansion is convergent due to the lack the critical point at zero temperature in $\tilde{Z}$ and $Z$. The structure of $r_n$ takes the form: $r_n=2^{A (1-n)}  \frac{\tilde{Z}}{Z}$, where
\begin{equation}
	\frac{\tilde{Z}}{Z}   =  \frac{    2^{A-1}2^{n-1}  \left[ 1+ \text{higher order corrections}  \right]}{  \left[  \text{ 1+ \text{higher order corrections}  }   \right]},
\end{equation}
where we notice the factor $ 2^{A-1}2^{n-1}  $ arising from the symmetry appears as a multiplicative factor in all orders of the expansion since different excitations related by the symmetry gives the same Boltzmann weight. Also, the corrections from higher order, which involve both connected and disconnected spin flips can be reorganized into an exponential of sum of only the connected spin flips as guaranteed by the linked cluster theorem. Since the cluster in expanding $\tilde{Z}$ involves three types of clusters: clusters only in the three dimensional bulk, clusters only on the n replicas of the two dimensional boundary, and the clusters connecting the three dimensional bulk and the two dimensional boundary, the linked cluster theorem implies that 

\begin{equation}
	\frac{\tilde{Z}}{Z}  =     2^{A-1}2^{n-1}    \frac{   e^{g_{\text{bulk}} (n,\beta    ) V  + g_{\partial}(n,\beta) A  }     }{  e^{g_{\text{bulk}}(n,\beta)V}}=2^{ A+n-2  } e^{ g_{\partial}(n,\beta) A       },
\end{equation}
where $g_{\text{bulk}}(n,\beta \to \infty)  =   g_{\partial}(n,\beta\to \infty) =0 $. Hence, $
r_n=2^{(A-1)(2-n)} e^{ g_{\partial}(n,\beta) A       }$, giving the Renyi negativity 
\begin{equation}
	R_n= \frac{1}{2-n} \log r_n = A \left( 1+\frac{1}{2-n}  g_{\partial}(n,\beta )  \right) \log 2   -\log 2.
\end{equation}
Thus the finite temperature excitations only alter the area-law coefficient, while the subleading term $\log 2$ survives until the breakdown of the perturbative series. Below we explicitly perform the expansion up to the order $O(e^{-24\beta  \lambda_B})$ as a demonstration. For the partition function $Z$, the lowest order of excitation is given by flipping four plaquette operators connected by a link contributing to order $O(e^{-8n\beta \lambda_B   })$ with degeneracy $3L^3$. For the  partition function $\tilde{Z}$, an elementary excitation is also given by flipping four plaquettes sharing a link. One can choose four plaquettes in the bulk, contributing to order $O(  e^{-8n\beta \lambda_B   }  )$ with degeneracy $3L^3-5A$, 3 plaquettes in the bulk plus 1 plaquette on the boundary giving $  O( e^{  -2 (3n+ 1)\beta \lambda_B } )$ with degeneracy $4A$, or four plaquettes on the boundary giving $O( e^{  -8\beta \lambda_B } )$ with degeneracy $A$. Note that when considering flipping the spins on the boundary replicas, one needs to perform the same spin flip on even number of replicas to respect the constraint $ \prod_i \delta\left(  \prod_{\alpha=1}^n\tau_i^{\alpha} =1   \right)$. To order $O(e^{-24\beta  \lambda_B})$ with Renyi index $n\geq 4$, it is sufficient to consider only the excitations on the boundary, and we find 

\begin{equation}
	\frac{\tilde{Z}}{Z}= 2^{A-1}2^{n-1} \left[ 1+A \binom{n}{2} e^{-16\beta\lambda_B }   +2A\binom{n}{2} e^{-24\beta\lambda_B } +\cdots \right]
\end{equation}
by considering the excitations as follows: choosing two replicas to perform a single spin flip in each replica at the same lattice site gives the correction of $O(e^{-16 \beta \lambda_B }) $ with degeneracy $A\binom{n}{2}$. The next order is to choose two replicas, and perform two neighboring spin flips on each replica, contributing order $O(e^{-24 \beta \lambda_B})$ with degeneracy $2A\binom{n}{2}$. Hence Renyi negativity reads

\begin{equation}
	R_n=A\left\{  \log 2 -\frac{1}{n-2} \left[   \binom{n}{2} e^{-16\beta\lambda_B}   +2 \binom{n}{2}  e^{-24\beta \lambda_B}  + \cdots  \right]      \right\}   -\log 2 .
\end{equation}
Beyond $O(e^{-24\beta  \lambda_B})$, one needs to consider the excitation by flipping four plaquettes involving bulk $B_p$ connected by a link. This can be four plaquettes in the bulk giving $O(e^{-8n\beta \lambda_B   })$, or 3 plaquettes in the bulk plus 1 plaquette on the boundary giving $O( e^{  -2 (3n+ 1)\beta \lambda_B } )$. Nevertheless, a crucial insight is that the dimension of the symmetry group $2^{A-1}2^{n-1}$ appears in all orders of the perturbative series, and hence the subleading $\log 2$ is expected to survives until the breakdown of the perturbative series.  

\begin{itemize}
	\item odd n
\end{itemize}
To gain intuition, we will first discuss zero temperature limit, and perform a low temperature expansion later. Similar to the even $n$ case, all the $B_p$ in $Z$ are set to be one, while for $\tilde{Z}$, only the $B_p$ in the bulk are set to one, giving
\begin{equation}
	r_n =  2^{A(1-n)} \sum_{   \{  B_{ij}\}}  \sum_{\{\tau^{1}_i \}  }  \cdots \sum_{\{\tau^{n}_i \}  } \prod_i \delta\left(  \prod_{\alpha=1}^n\tau_i^{\alpha} =1   \right)   \prod_{\alpha=1}^n\prod_{\expval{ij}} \delta\left( B_{ij}\tau^{\alpha}_i\tau^{\alpha}_j=1\right).
\end{equation}
Similar to the case of even $n$, there exists a global spin flip symmetry for each replica subject to the constraint $\prod_{i} \delta\left( \prod_{\alpha=1}^n \tau_i^{\alpha } =1 \right)$, hence giving $2^{n-1}$ for the dimension of the symmetry group. On the other hand, the local transformation by sending $B_{ij} \to -B_{ij}$ for four links emanating from a site $i$ accompanied by $\tau^{\alpha}_{i} \to -\tau^{\alpha}_i ~\forall \alpha$ does not exist for odd $n$. This is because taking $\prod_{\alpha=1}^n \tau_i^{\alpha} \to   - \prod_{\alpha=1}^n  \tau_i^{\alpha} $ for odd $n$ violates the constraint $\delta\left(  \prod_{\alpha=1}^{n}   \tau_i^{\alpha} =1\right) $. As a result, 
\begin{equation}
	r_n=2^{-nA }2^{A}  2^{n-1}  =2^{(A-1)(1-n)}.
\end{equation}
Correspondingly, the Renyi negativity is $R_n= \frac{1}{1-n} \log r_n= A\log 2-\log 2$. At finite temperature, to the leading order, we neglect the excitation for the bulk $B_p$, so the boundary plaquette $B_{ij}$ remains frustration free. The leading order correction is given by four violations of $ \delta \left(  B_{ij} \tau_i\tau_j =1\right)$ on bonds emanating from a single site, contributing to order $O(e^{-8\beta \lambda_B})$ with degeneracy $nA$. Notice that we cannot perform a single spin flip to achieve this since the constraint among different replicas $\delta\left(  \prod_{\alpha=1}^n\tau_i^{\alpha} =1   \right) $ will be violated. However, the violation of four bonds can be performed by sending $B_{ij} \to -B_{ij}$ for four $B_{ij}$ sharing the same lattice site $i$, and sending $\tau_i^{\alpha}\to -\tau_i^{\alpha} $ for only $n-1$ repicas, creating exactly the excitations of four bonds in only one replica. This implies that when performing the low temperature expansion starting from one of the $2^{n-1}$ ground states, effectively there is no constraint on the allowed spin filps. Hence we consider the  excitations as follows: a single spin flip of $O(e^{-8\beta \lambda_B })$ with degeneracy $nA$, two neighboring spin flips on the same replica of $O(e^{-12\beta \lambda_B })$ with degeneracy $2nA$. For $O(e^{-16\beta\lambda_B})$, the excitations are two non-neighboring spin flips on the same replica with degeneracy $n\left(  \binom{A}{2}-2A \right)$, two spins on two different replicas of $O(e^{-16\beta \lambda_B})$ with degeneracy $  \binom{n}{2}A^2$, four spins at the vertices of a square unit cell on the same replica with degeneracy $nA$, three connected spin flips in a row/column on the same replica with degeneracy $2nA$, and three connected spin flips at the vertices of a triangle on the same replica with degeneracy $4nA$. Therefore,

\begin{equation}
	r_n=2^{A(1-n)} 2^{n-1}  \left\{    1+nA e^{-8\beta \lambda_B} +2nA e^{-12 \beta \lambda_B}      +  e^{-16 \beta\lambda_B}  \left[        n \left(  \binom{A}{2}  -2A \right) +    \binom{n}{2}  A^2  + nA+2nA+ 4nA   \right]     +\cdots\right\}.
\end{equation}
Up to order $O(e^{-16\beta \lambda_B})$, odd Renyi negativity $R_n=\frac{1}{1-n} \log r_n$ reads 
\begin{equation}
	R_n =\frac{1}{1-n} \log r_n =  A   \left[  \log 2 -\frac{n}{n-1} \left( e^{-8\beta \lambda_B}  +2e^{-12\beta \lambda_B}  +  \frac{9}{2}  e^{-16\beta \lambda_B}  \right) \right] -\log 2,
\end{equation}
where we see $O(A^2)$ term in $r_n$ arising from the disconnected excitations cancels out after taking logarithm as demanded by the linked cluster theorem.

\section{4D toric code}
We consider the four dimensional toric code with the Hamiltonian: 
\begin{equation}
	H_T=-\lambda_A \sum_l A_l -\lambda_B \sum_{c}B_c,
\end{equation}
Here $l$ and $c$ label an link (1-cell) and a cube (3-cell) respectively. Spins reside on each face (2-cell), and $A_l$ is the product of 6 Pauli-X operators on the faces adjacent to the link $l$, $B_c$ is the product of 6 Pauli-Z operators on the faces around the boundary of the cube $c$. 

\subsection{General expression of (Renyi) negativity}
We define $\hat{x}, \hat{y}, \hat{z}, \hat{t}$ for four spatial directions, and choose $t=0$ as the location of the bipartition surface. This bipartition boundary is a three dimensional cubic lattice since it is the projection of a four dimensional lattice. The boundary operators invove $A_l$ operators, which live on the link of the cubic lattice, and $B_c$ operators, which live on the face of the three dimensional lattice as they are the three dimensional cube by dragging the face along $\hat{t}$ direction. Note that we replace those boundary $B_c$ operators by $B_f$ below. Write $H$ as $H_\mathcal{R}+H_{\overline{\mathcal{R}}}+H_{\mathcal{R}\overline{\mathcal{R}}}$, we first focus on the boundary part of the density matrix $
e^{-\beta H_{\mathcal{R}\overline{\mathcal{R}}}} = e^{  \beta \lambda_A \sum^{\partial}_{l} A_l + \beta \lambda_f \sum^{\partial}_{f} B_f }$ where $\partial$ on the summation symbol means summing only the boundary operators. Similar to the calculation in 2D and 3D, we write $e^{-\beta H_{\mathcal{R}\overline{\mathcal{R}}}}$ as a summation involving $n_l$, $\sigma_f$ variables for denoting whether a star, and a plaquette term is present or not. A partial transpose gives 
\begin{equation}
	\left[\prod_{l}^{\partial} A_l^{n_l } \prod_{f}^{\partial } B_f^{\sigma_f}\right]^{T_{\partial \overline{\mathcal{R}}}}  = \prod_{l}^{\partial} A_l ^{n_l } \prod_{f}^{\partial } \left(B_f  \prod_{l\in \partial f} \tau_{l}  \right)^{\sigma_f} ,
\end{equation}
where $\tau_l=1-2n_l=\pm 1$. Then one finds,

\begin{equation}\label{append:4d_pt}
	\left( e^{-\beta H_{\mathcal{R}\overline{\mathcal{R}}}   }   \right)^{T_{\partial \overline{\mathcal{R}}}}  = \cosh^{N_l}(\beta\lambda_A)  \sum_{  \{\tau_l  \} }  e^{\sum_{l}^{\partial } \frac{1-\tau_l}{2} \log \left(  A_l \tanh(\beta \lambda_A)     \right)  +   \beta \lambda_B \sum^{\partial}_ {  f }  B_f \prod_{l \in \partial f} \tau_{l}  },
\end{equation}
where $N_l=3L^3$ is the number of links of a three dimensional cube. Basically this corresponds to a partition function for a 3D cubic lattice, where Ising spin $\tau_l$ lives on the link, subjected to on-site field and four-spins interaction around each face. Therefore, by writing the Renyi negativity as $R_n=b_n \log r_n$, we finds 

\begin{equation}\label{append:4d_rn}
	r_n= \frac{   \sum_{\{A_l\}} \sum_{\{ B_c \}} f( \{A_l\}, \{B_c  \}    )  e^{    -n\beta(H_\mathcal{R}+H_{\overline{\mathcal{R}}}) }     \left[  \left(e^{-\beta H_{\mathcal{R}\overline{\mathcal{R}}}}\right) ^{T_{\partial \overline{\mathcal{R}}}}\right]^n   }{       \sum_{\{A_l\}} \sum_{\{ B_c \}}  f( \{A_l\}, \{B_c  \}    ) e^{    -n\beta(H_\mathcal{R}+H_{\overline{\mathcal{R}}}+H_{\mathcal{R}\overline{\mathcal{R}}}) } },
\end{equation}
where $f( \{A_l\}, \{B_c  \}    )$ imposes the local constraint
\begin{equation}\label{append:four_dim_constraint}
	f( \{A_l\}, \{B_c  \}    )   = \prod_{ 4\text{-cell}  }\delta \left( \prod_{c\in \partial 4\text{-cell}}  B_c =1 \right) \prod_{0\text{-cell}}  \delta  \left(   \prod_{0\text{-cell} \in\partial l}    A_l =1   \right).
\end{equation}
Note that there can be global constraints as well, depending on the imposed boundary conditions.

\subsection{Negativity at $T=0$}
At $T=0$, imposing open boundary condition along $\hat{t}$ spatial direction and periodic boundary condition along $\hat{x}, \hat{y}$, and $\hat{z}$ direction, we separate a subsystem $\mathcal{R}$ from its complement $\overline{\mathcal{R}}$ using a three dimensional boundary (a three dimensional lattice). Define $N_{\partial }$ for the number of boundary $A_l$ operator, one finds 
\begin{equation}
	\norm{\rho^{T_{\overline{\mathcal{R}}}}}_1=  2^{-N_{\partial } }  \sum^{\partial }_{\{A_l\}} \sum^{\partial}_{\{ B_f \}} f_{\partial }( \{A_l\}, \{B_f  \}    )  \prod_{\mu} \delta \left(     \prod_{f\in \mu } B_f=1   \right)   \abs{  \sum_{\{\tau   \}  }   \prod_l  A_l^{\frac{1-\tau_{l}}{2}}   \prod_{  \text{face}    }   \delta\left(   B_f  \prod_{l\in \partial \text{face}}      \tau_l\right)   } ,  
\end{equation}
where the constraint $\delta\left(   B_f  \prod_{l\in \partial \text{face}}      \tau_l\right)$ arises to maximize the boltzmann weight while the constraint $ \prod_{\mu} \delta \left(     \prod_{f\in \mu } B_f=1   \right)  $ with $\mu$ labeling $xy, yz$, or $zx$ planes arises so that each local term $B_f  \prod_{ l\in  \text{face} }  \tau_l$ can be independently maximized (frustration free condition). On the other hand, setting all the bulk operators by one in the constraint (Eq.\ref{append:four_dim_constraint}) gives the constraint $ f_{\partial }( \{A_l\}, \{B_f  \}    ) $ imposed on the boundary operators:
\begin{equation}\label{append:4d_constraint}
	\prod_{\text{vertex} }  \delta\left( \prod_{l\in \text{vertex }  } A_l=1 \right)\prod_{ \text{cube}   }     \delta\left( \prod_{f\in \partial \text{cube}    }B_f=1 \right),
\end{equation}
where it gives $L^3-1$ number of constraints for $A_l$ and $B_f$ respectively. To calculate the argument in the absolute value sign, one can start from a reference $\{\tau\}$ configuration satisfying the flux constraint determined by $B_f$, and the gauge transformation $\tau_l\to -\tau_l$ for all links connected to a vertex can generate a new allowed $\{ \tau_l \}$ configurations. Note that different $\tau_{l}$ configurations have the same sign by noticing that $\prod_lA_l^{\frac{1-\tau_{l}}{2}}$ is invariant under the gauge transformation due to the constraint Eq.\ref{append:4d_constraint}. Meanwhile, one can perform a transformation $\tau_l\to -\tau_l $ for all the links along $\hat{x}$, $\hat{y}$, or $\hat{z}$ directions. Note that this operation respects the flux constraint, but cannot be generated by the local constraint acting on the vertices. Since there are $L^3-1$ number of independent generators  for the local gauge transformation, one finds

\begin{equation}
	\begin{split}
		\abs{  \sum_{\{\tau   \}  }   \prod_l  A_l^{\frac{1-\tau_{l}}{2}}   \prod_{  \text{face}    }   \delta\left(   B_f  \prod_{l\in \partial \text{face}}      \tau_l\right)   }&=  2^{L^3-1}\abs{    \prod_l  A_l^{\frac{1-\tau_{l}}{2}}  \left( 1+\prod_{l\in \text{x-\text{link}}} A_l     \right)     \left( 1+\prod_{l\in \text{y-\text{link}}} A_l     \right)    \left( 1+\prod_{l\in \text{z-\text{link}}} A_l     \right) }\\
		&= 2^{L^3-1} 2^3 \prod_{ \mu=x,y,z   } \delta\left(    \prod_{l\in \mu-\text{link}} A_l =1 \right).
	\end{split}
\end{equation}
Finally using the ingredient above, one finds $\norm{\rho^{T_{\overline{\mathcal{R}}}}}_1 =2^{\text{number of independent } B_f  }= 2^{N_{\partial }-(L^3-1+3)}= 2^{2L^3-2}$,
which gives the negativity 
\begin{equation}
	E_N =  \left(   L^3-1\right)2 \log2.
\end{equation} 
Notice that the area law coefficient and the subleading term are both $2\log2$ as opposed to $\log2$ as in the 2D and 3D toric code. A similar calculation shows that Renyi negativity follows the same expression as well.

\subsection{Low temperature expansion of Renyi negativity}
Using Eq.\ref{append:4d_rn}, we perform a perturbative calculation at low temperature up to $O(e^{-24\beta\lambda_B})$ and $O(e^{-24\beta\lambda_B})$  for even Renyi negativity. For simplicity, we will impose periodic boundary condition on all spatial directions, and consider $t=t_1$, $t=t_2$ as two bipartition surfaces to divide the total system into $\mathcal{R}$ and its complement $\overline{\mathcal{R}}$. Similar to the discussion to the perturbative calculation in 3D toric code, it suffices to calculate the negativity contribution from one of the bipartition surface since both surfaces contributes equally, and independently to the total negativity. Up to the order we consider, in Eq.\ref{append:4d_rn}, we can choose all operators to be one in the denominator. In the numerator, the leading order excitations only occur on the boundary so that we can set the bulk operators by one. Finally, define $N=3L^3$, one finds the negativity contribution from one bipartition boundary reads
\begin{equation}\label{append:4d_expansion}
	r_n=2^{-nN}  e^{-nN\beta \lambda_B}  \sum_{\{  A_l \}} \sum_{ \{ B_f \}  }f_{\partial} (  \{A_l    \}  ,\{ B_f  \})   \left[   Z\left(\{ A_l  \}   ,\{  B_f  \}  \right)   \right]^n,
\end{equation}
where 
\begin{equation}
	Z\left(\{ A_l  \}   ,\{  B_f  \}  \right)  =  \sum_{\{\tau_{l}\}}  \left(    \prod_{l} A_l^{\frac{1-\tau_{l}}{2}}  \left(   1+\tau_{l}e^{-2\beta\lambda_A}   \right)  \right)   e^{\beta \lambda_B \sum_f B_f\prod_{l\in \partial f} \tau_l  },
\end{equation}
and 
\begin{equation}
	f_{\partial} (  \{A_l    \}  ,\{ B_f  \})=\prod_{\text{vertex} }  \delta\left( \prod_{l\in \text{vertex }  } A_l=1 \right)\prod_{ \text{cube}   }     \delta\left( \prod_{f\in \partial \text{cube}    }B_f=1 \right).
\end{equation}
Similar to the calculation at $T=0$, the summation over $\tau_l$ can be written as summation over all possible gauge transformations. The excitation associated with $\lambda_A$ can be obtained by expanding $\prod_{l} \left(   1+\tau_l e^{-2\beta \lambda_A}  \right)$. Note that only the gauge-invariant terms can survive after summing the gauge group so for example, the leading order excitation is given by the product of four $\tau_{l}$ on a plaquette, giving the boltzmann factor $e^{-   8\beta\lambda_A}$, and the next excitation is the product of six $\tau_l$ spanning two nearest neighboring plaquettes. On the other hand, the excitation associated with $\lambda_B$ can be obtained by flipping a $\tau_{l}$ spin (creating four plaquette excitations), giving the boltzmann factor  $e^{-   8\beta\lambda_B}$. By considering the excitation by flipping spins and expanding $\prod_{l} \left(   1+\tau_l e^{-2\beta \lambda_A}  \right)$, one finds

\begin{equation}\label{append:4d_z}
	\begin{split}
		Z\left(\{ A_l  \}   ,\{  B_f  \}  \right) 
		= & 2^{L^3-1}e^{N\beta \lambda_B} \left[\prod_{l}A_l^{\frac{1-\tau_l}{2}} \right]\left[   \prod_{\mu=x,y,z}\left(  1+\prod_{ l\in  \mu-\text{link} }  A_l \right)   \right] \\
		&\left(    1+ e^{-8\beta\lambda_A} T^A_1+ e^{-12 \beta\lambda_A} T^A_{2} + O(e^{-16\beta \lambda_B })  \right)   \left(    1+ e^{-8\beta\lambda_B} T^B_1+ e^{-12 \beta\lambda_B} T^B_{2} + O(e^{-16\beta \lambda_B })  \right).
	\end{split}
\end{equation}
Here $T_1^B=\sum_f B_f$, $T_2^B=\sum_{\expval{f,f'}} B_fB_f'$ with $\expval{ f,f' }$ denoting two neighboring plaquettes while $T^A_1= \sum_{l}A_l $, $T^A_2 =\sum_{   \expval{l,l'}} A_lA_{l'}$ with $\expval{ l,l' }$ denotes two edges of a plaquette. Note the number of $\expval{f,f'}$ (or$\expval{l,l'   }$) is $12L^3$. Technically, Eq.\ref{append:4d_z} is not correct since when flipping a $\tau_l$ spin to create excitations in $\lambda_B$ (corresponding to $T_1^{B}$), the product of four $\tau_{s}$ spins will be $-B_f$ instead of $B_f$ if the product involves the flipped spin in $T_1^A$. Nevertheless, one can check the sign does not affect the result for even Renyi negativity. For even $n$, one finds

\begin{equation}
	\begin{split}
		Z^n\left(\{ A_l  \}   ,\{  B_f  \}  \right) 
		= & 2^{n(L^3+2)}e^{nN\beta \lambda_B}      \prod_{\mu=x,y,z}  \delta\left( \prod_{l\in \mu-\text{link}}A_l  =1 \right)   \\
		&\left(    1+ e^{-8\beta\lambda_A} T^A_1+ e^{-12 \beta\lambda_A} T^A_{2} + O(e^{-16\beta \lambda_A })  \right)^n   \left(    1+ e^{-8\beta\lambda_B} T^B_1+ e^{-12 \beta\lambda_B} T^B_{2} + O(e^{-16\beta \lambda_B }) \right)^n.
	\end{split}
\end{equation}
Notice that $T_{1(2)}^{A(B)}$ under summing $A_l$, $B_f$ vanish due to symmetry, it suffices to consider the correction term such as $\left(  T_{1(2)}^{A(B)}  \right)^2$. For example, consider $( T_1^{B})^2 = \sum_{f,f'} B_fB_{f'}$, when summing over $B_{f}$ (respect to the constraint for $B_f$), only the diagonal part ($f=f'$) survives (this explains the aforementioned statement that writing $B_f$ or $-B_f$ does not matter). Therefore, one just need to count the number of term in $T_1^{A}$, and similarly for $T_{1(2)}^{A(B)}$.  Finally, one finds 
\begin{equation}
	r_n=2^{(2-n)  (2L^3-2)   }  \left( 1+3L^3 \binom{n}{2}  e^{-16\beta \lambda_A}  + 12L^3\binom{n}{2} e^{-24\beta \lambda_A}      \right)  \left( 1+3L^3 \binom{n}{2}  e^{-16\beta \lambda_B}  + 12L^3\binom{n}{2} e^{-24\beta \lambda_B}      \right).
\end{equation}
It follows that the \textit{even} Renyi negativity contribution from a single bipartition surface is 
\begin{equation}
	R_n=L^3 \left[    2\log 2 -\frac{3n(n-1)}{2(n-2)}\left(  e^{-16\beta \lambda_A} +e^{-16\beta \lambda_B}  \right)    -\frac{6n(n-1)}{(n-2)}\left(  e^{-24\beta \lambda_A} +e^{-24\beta \lambda_B}  \right)     \right]  -2\log 2.
\end{equation}
Therefore, increasing temperature only decreases the area- law coefficient of Renyi negativity without changing the subleading term in this low temperature regime.

\section{Calculation of negativity using tensor network representation }
Here we present an alternative derivation for the partial transposed density matrix and the negativity using the tensor network representation. 
\subsection{2D toric code}
\begin{figure}
	\centering
	\begin{subfigure}{0.40\textwidth}
		\includegraphics[width=\textwidth]{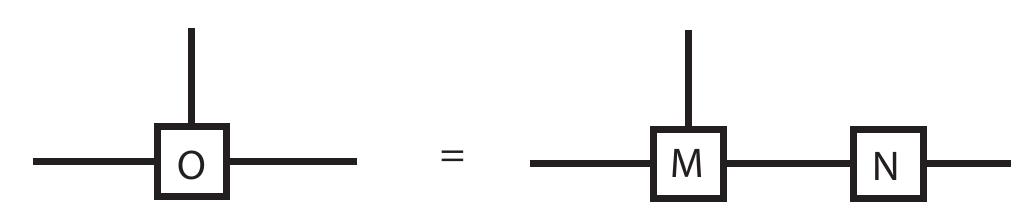}
	\end{subfigure}
	\hspace{2cm}
	\begin{subfigure}{0.40\textwidth}
		\includegraphics[width=\textwidth]{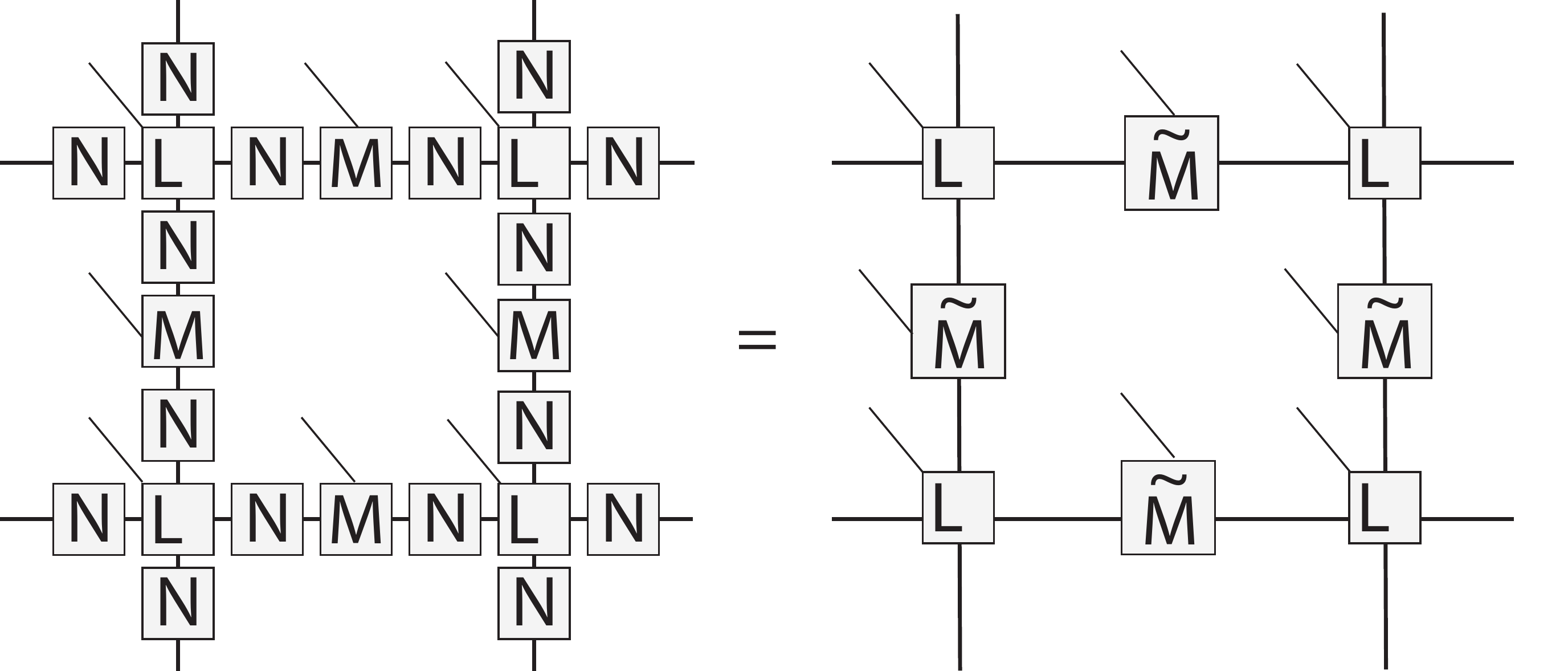}
		\label{peps}
	\end{subfigure}
	\caption{Left: the building block of the MPS for $\psi$ in 2D toric code. Right: the building block of the tensor network for computing negativity in 3D toric code as $\lambda_B \rightarrow \infty$.   }  
	\label{fig:network}  
\end{figure}

We consider the 2d toric code model with open boundary conditions, in which all the stabilizers are independent variables.  The thermal density matrix is $
\rho = \frac{1}{2^N}\prod_s (I+\tanh(\beta \lambda_A)A_s) \prod_p (I+\tanh(\beta \lambda_B) B_p) $, and for convenience we define $t_A \equiv \tanh(\beta \lambda_A), t_B \equiv \tanh(\beta \lambda_B)$. In calculating the negativity, the partial transpose only affects alternating star and plaquette operators along the boundary of the partition; we label these operators from $1$ to $2L$, where $L$ is the physical length of the boundary.  Every time when two adjacent operators appear in the expansion of the density matrix, the corresponding term picks up a negative sign as a result of the partial transpose.  Because the bulk stabilizer values are independent, we factor $\rho^{T_A} = \rho_{bulk} \otimes \rho_{bd}$, and $\rho_{bulk}$ is not affected by the partial transpose. The eigenvalue of $\rho_{bd}$ associated to a given configuration of boundary stabilizer values is
\beq
e(\{A_s, B_p\}) = \frac{1}{2^{2L}} \sum_{\vec{x} = (x_1,...x_{2L})} \psi(\vec{x}) \prod_s (t_A A_s)^{x_s} \prod_p (t_B B_p)^{x_p}  \nonumber
\eeq 
where each $\vec{x}$ is a set of $1$s and $0$s respectively indicating whether a given operator appears in the (boundary-restricted) expansion of the density matrix or not; $\psi$ is the function which assigns negative signs anytime two adjacent operators both appear. Matrix product states (MPS) provide a convenient way to evaluate this quantity.  First notice that $\psi$ can be built from a simple MPS representation: on every site, $M^1 =
\begin{pmatrix}
1 & 0 \\
0 & 0
\end{pmatrix} 
$, $M^0 =
\begin{pmatrix}
0 & 0 \\
0 & 1
\end{pmatrix}$
simply convey the physical value $x=1 (0)$ to the virtual level.  
Between sites, $N =
\begin{pmatrix}
-1 & 1 \\
1 & 1
\end{pmatrix}$
implements the negative sign in the case that adjacent physical sites are both $1$. These can be contracted to yield an MPS with $
O^1 =
\begin{pmatrix}
-1 & 1 \\
0 & 0
\end{pmatrix}
$, $
O^0 =
\begin{pmatrix}
0 & 0 \\
1 & 1
\end{pmatrix}
$ (see Fig.\ref{fig:network}). To get $e(\{C_j\})$, where $C_j$ denotes either stabilizer, we first build in $t_j C_j$ by making the modification $O_j^1 =
\begin{pmatrix}
-t_j C_j & t_j C_j \\
0 & 0 
\end{pmatrix}
$
and we sum over all operator strings (all $x$ configurations) by simply adding $O^0,O_j^1$ on every site before taking the trace of the matrix product.  Hence, 
\beq
e(\{s_j\}) = \frac{1}{2^{2L}} \tr \left[  \prod_{j=1}^{2L} \begin{pmatrix}
	-t_j C_j & t_j C_j \\
	1 & 1 
\end{pmatrix}\right],
\eeq
and the negativity (with the bulk piece subtracted) is $E = \log \left[  \sum_{\{C_j\}} |e(\{C_j\})|  \right]$. In fact, we can directly relate this approach to transfer matrix introduced in Eq.\ref{append:transfer} by taking the product of two adjacent matrices (with two open physical indices $C_A, C_B$ which will correspond to the $A, B$ stabilizer variables).  The resulting matrix is $
\begin{pmatrix}
t_A C_A (1+t_B C_B) & t_A C_A (1-t_B C_B) \\
1-t_B C_B  & 1+t_B C_B 
\end{pmatrix}$, which is exactly the transfer matrix in Eq.\ref{append:transfer} multiplied by $1/\cosh(\beta \lambda_B)$.

\subsection{Equivalence of negativity in 2D/3D toric codes for $\lambda_B \rightarrow \infty$ limit}

For $\lambda_B \rightarrow \infty$, $t_B = 1$ and it is more convenient to absorb the $N$ tensors into the $M$ tensors of the plaquette stabilizers.  It is straightforward to check that
\beq
e(\{C_A\},\{C_B\})=\frac{1}{2^{L}} \tr\left[ \prod_j M^{C_{A,j}} {\tilde M}^{C_{B,j}}  \right],
\eeq 
where $
M^{C_{A,j}}=\begin{pmatrix}
t_A C_{A,j} & 0 \\
0 & 1 
\end{pmatrix},
{\tilde M}^{C_{B,j}=1}=\begin{pmatrix}
1 & 0 \\
0 & 1 
\end{pmatrix},
{\tilde M}^{C_{B,j}=-1}=\begin{pmatrix}
0 & 1 \\
1 & 0 
\end{pmatrix}.$ Thus, once a configuration $\{C_B\}$ is chosen, the indices on the $C_A$ sites (essentially, whether the site contributes $1$ or $t_A C_{A,j}$) are fixed up to a global $Z_2$ operation.  

The negativity calculation for 3D toric code in the $\lambda_B \rightarrow \infty$ limit has the same structure.  In this limit, one does not need to worry about the constraint that the product of six plaquette operators on the faces of a cube is identity.  Hence, the calculation proceeds in analogy with the 2D case; it localizes to the two-dimensional boundary specified by the negativity partition, and $e(\{C_A\},\{C_B\})$ is specified by the 2D tensor network shown in Fig.\ref{fig:network}.  In the figure, $M, N$ are the same tensors as in the 2D toric code case, and the $L$ tensor is defined by $L^{C_A}_{\alpha \beta \gamma \rho} = t_A C_A \delta_{\alpha 0} \delta_{\beta 0} \delta_{\gamma 0} \delta_{\rho 0} +  \delta_{\alpha 1} \delta_{\beta 1} \delta_{\gamma 1} \delta_{\rho 1}$.  Here $C_A$ is the physical index and $\alpha, \beta, \gamma, \rho$ are the four virtual indices. 

Again, it is convenient to contract the $N, M$ tensors into ${\tilde M}$ tensors.  Moreover, once the configuration $\{C_B\}$ is chosen, the indices on the $C_A$ sites are fixed up to a global $Z_2$.  Thus, the negativity is exactly the same as for the 2D toric code case, with number of $C_A$ sites now $L_x L_y$ as opposed to $L_x$.

\section{Sign structure of the partial transposed density matrix}
Here we show that the sign of eigenvalues of $\rho^{T_{\overline{\mathcal{R}}}}$ at zero temperature is determined from the parity of braids between two types of operators in toric code models. To begin with, since $\rho^{T_{\overline{\mathcal{R}}}} =\frac{1}{Z} e^{-\beta (H_\mathcal{R}+H_{\overline{\mathcal{R}}})} \left(   e^{-\beta H_{\mathcal{R}\overline{\mathcal{R}}}} \right)^{T_{\partial \overline{\mathcal{R}}}}$, it is sufficient to consider only on the sign of eigenvalues of $\left(   e^{-\beta H_{\mathcal{R}\overline{\mathcal{R}}}} \right)^{T_{\partial \overline{\mathcal{R}}}}$, which determines the sign of eigenvalues of $\rho^{T_{\overline{\mathcal{R}}}}$. In the 2D toric code,

\begin{equation}
	\left(   e^{-\beta H_{\mathcal{R}\overline{\mathcal{R}}}} \right)^{T_{\partial \overline{\mathcal{R}}}}     \sim \sum_{\boldsymbol{\tau } }  \left[ e^{\beta \lambda_B \sum_i B_i \tau_{i}  \tau_{i+1}  }  \prod_{i=1}^L  \left( A_i \tanh(\beta \lambda_A)   \right)^{\frac{1-\tau_{i}}{2}}  \right],
\end{equation}
As $\beta \to \infty$, one only needs to consider the frustration free configurations $B_i\tau_i \tau_{i+1}=1$. Therefore, once $\tau_1$ is determined, the rest of the spins follows $\tau_i= \tau_1B_1 B_2 \cdots B_{i-1}$, and

\begin{equation}
	\left(   e^{-\beta H_{\mathcal{R}\overline{\mathcal{R}}}} \right)^{T_{\partial \overline{\mathcal{R}}}}   \sim  \delta\left( \prod_{i=1}^L{B_i}=1     \right)  \left[  A_1^{\frac{1-\tau_1}{2}}  A_2^{\frac{1-\tau_1B_1}{2}}  \cdots A_L^{\frac{1-\tau_1  B_1 B_2\cdots B_{L-1} }{2}  }  +    A_1^{\frac{1+\tau_1}{2}}  A_2^{\frac{1+\tau_1B_1}{2}}  \cdots A_L^{\frac{1+\tau_1  B_1 B_2\cdots B_{L-1} }{2}  }      \right]    .
\end{equation}
There are two terms since $\tau_1$ can be 1 or -1. Also note that the global constraint for $B_i$ is necessary for frustration free condition. The above equation can be further simplified as

\begin{figure}
	\centering
	\includegraphics[width=0.4\textwidth]{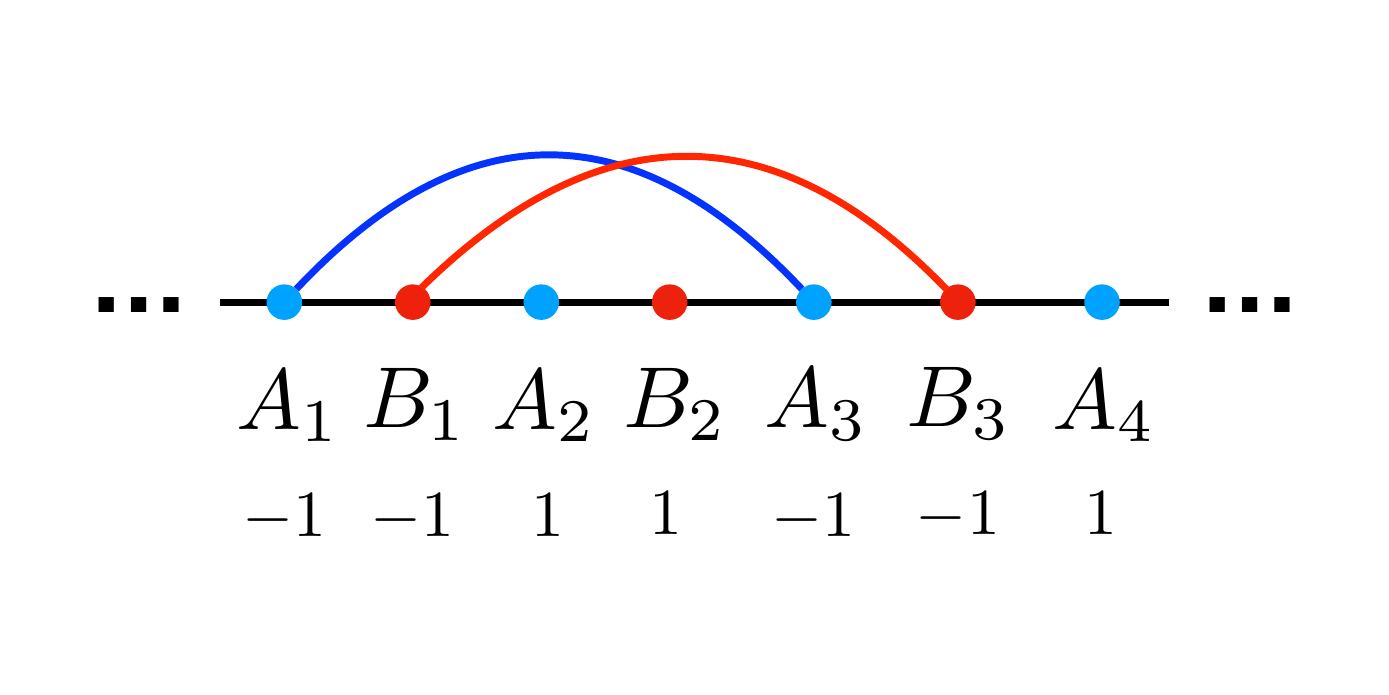}
	\caption{On the bipartition boundary in the 2D toric code, assign $\pm 1$ for each boundary operators, and connect $A_i,A_j$/ $B_i,B_j$ with a string (A-string/B-string) if $A_i=A_j=-1$/  $B_i=B_j=-1$. Odd/even total number of crossing between A-string and B-string gives negative/positive eigenvalues of $\left( e^{-\beta H}  \right)^{T_{\overline{\mathcal{R}}}}$.     }
	\label{fig:2dbraiding}
\end{figure}

\begin{equation}
	\left(   e^{-\beta H_{\mathcal{R}\overline{\mathcal{R}}}} \right)^{T_{\partial \overline{\mathcal{R}}}}  \sim \delta\left( \prod_{i=1}^L A_i=1     \right) \delta\left( \prod_{i=1}^L{B_i}=1     \right)  \prod_{i=2}^L A_i^{ \frac{1}{2} \left(   1- B_1B_2 \cdots B_{i-1}  \right) }.
\end{equation}
Specify the boundary operators in the order of $A_1, B_1, A_2, B_2, \cdots $, and assign 1 or -1 for the operators, we connect $A_i$ and $A_j$ (or $B_i$ and $B_j$) by an string extending into the bulk if $A_i=A_j=-1$ (or $B_i=B_j=-1$). The eigenvalue of $\left(   e^{-\beta H_{\mathcal{R}\overline{\mathcal{R}}}} \right)^{T_{\partial \overline{\mathcal{R}}}}  $ is negative (positive) when there is odd  (even) number of crossing between $A$-string and $B$-string (see Fig.\ref{fig:2dbraiding}). We make several remarks here. First, given multiple $A_i$ (or $B_i$) equal $-1$, the number of crossing is independent of how we connect operators. Second, the string is always closed since the two global constriants makes sure the sign flip of the operator always comes in pair. Third, at any finite temperature, the constraints disappears (assuming $\lambda_A, \lambda_B$ is finite), corresponding to the break of closed strings into open strings. Relatedly, the sign structure in terms of the crossing between two types of strings is destroyed.

One can apply a similar analysis to 3D and 4D toric code model. In the 3D toric code, connecting $A_s$ operators of minus sign gives a $A$-string while connect $B_p$ operators of minus sign gives a $B$-membrane. The eigenvalue of $ \left(   e^{-\beta H_{\mathcal{R}\overline{\mathcal{R}}}} \right)^{T_{\partial \overline{\mathcal{R}}}}  $ is negative ( positive ) when there is odd  (even) number of crossing between $A$-string and $B$-membrane. In the 4D toric code, connecting $A_l$ operators of minus sign gives a $A$-membrane while connect $B_c$ operators of minus sign gives a $B$-membrane. The eigenvalue of $\left(   e^{-\beta H_{\mathcal{R}\overline{\mathcal{R}}}} \right)^{T_{\partial \overline{\mathcal{R}}}}  $ is negative ( positive ) when there is odd  (even) number of crossing between $A$-membrane and $B$-membrane.

 \end{document}